\documentclass[fleqn,usenatbib,useAMS]{mnras}

\usepackage{graphicx, natbib, color, bm, amsmath, epsfig,hyperref}

\usepackage[T1]{fontenc}
\usepackage{ae,aecompl}
\usepackage{newtxtext,newtxmath}

\usepackage[normalem]{ulem}

\def\sss{\ensuremath{S_{2,L}(r)}}
\def\rhat{\ensuremath{\hat{\rvec}}}
\def\sigmavl{\ensuremath{S_{2,L,c_{i}}}}

\usepackage{graphicx, natbib, color, bm, amsmath, epsfig}

\def\sima{\ensuremath{sim1}}
\def\simb{\ensuremath{sim2}}
\def\simc{\ensuremath{sim3}}

\newcommand\jcompphys{J.~Comput.~Phys}

\newcommand{\msun}{\ensuremath{M_\odot}}

\newcommand{\tff}{\ensuremath{t_{\rm{ff}}}}

\definecolor{orange}{rgb}{1.        ,  0.54,  0}

\definecolor{meta}{rgb}{0.371,0.617,0.625} 

\newcommand{\dc}[1]{}

\def\vvec{\ensuremath{{\bf v}}}

\def\rvec{\ensuremath{{\bf r}}}
\def\xvec{\ensuremath{{\bf x}}}

\definecolor{pink}{rgb}{1.        ,  0.75294118,  0.79607843}
\definecolor{maroon}{rgb}{0.69019608,  0.18823529,  0.37647059}

\def\nn{\nonumber}

\definecolor{gray}{rgb}{0.5,0.5,0.5}

\newcommand{\percc}{\ensuremath{\rm{cm}^{-3}}}

\def\muG{\ensuremath{\mu\rm{G}}}

\def\hw{0.49}

\def\sigmal{\ensuremath{\sigma_{s}}}
\def\sigmavone{\ensuremath{\sigma_{v,1d}}}
\def\rhomax{\rho_{\rm{max}}}

\def\papertwo{Paper II}

\title{Collapsing Molecular Clouds with Tracer Particles: Part I, What
Collapses?}
\author[Collins, D. C. et al]{Collins, David C.$^{1}$ \thanks{email:
dccollins@fsu.edu}, Le, Dan$^{1}$, Jimenez
Vela, Luz$^{1}$
\\
$^{1}$Department of Physics, Florida State University, Tallahassee, FL}

\begin{document}
\maketitle

\begin{abstract}
To understand the formation of stars from clouds of molecular gas, one essentially needs to know two things:
What gas collapses, and how long it takes to do so.  We address these questions
by embedding pseudo-Lagrangian tracer particles in three simulations of
self-gravitating turbulence.  We identify prestellar cores at the end of the
collapse, and use the tracer particles to rewind the simulations to identify the
preimage gas for each core at the beginning of each simulation.  This is the
first of a series of papers, wherein we present the technique and examine
the first question: What gas collapses?
For the preimage gas at the t=0, we examine a number of quantities; the probability
distribution function (PDF) for several quantities, the structure function for
velocity, several length scales, the volume filling fraction, the overlap
between different preimages,  and fractal dimension of the preimage
gas.
 Analytic descriptions are found for the PDFs of density and
velocity for the preimage gas.  
We find that the preimage of a core is large and
sparse, and we show that gas for one core comes from many turbulent density
fluctuations and a few velocity fluctuations.  
We examine the convex hull for each preimage and show that gas
from different cores begins the simulation spatially intertwined with one
another in a fractal manner.
We find that binary systems have preimages that overlap in a fractal manner.
Finally, we use the density
distribution to derive a novel prediction of the star formation rate.  
\end{abstract}




%

\section{Introduction}
\label{sec.intro}

Stars form out of clouds of molecular gas that is barely held together by
gravity \citep{Dobbs11}.   They are turbulent, magnetized, self-gravitating, and once stars begin
to form the clouds are blown apart by radiation and outflows \citep{Chevance22}.  The extraordinary
violence of the supersonic flows within the clouds create overdensities that are massive
enough to decouple from the turbulence and
collapse \citep{Krumholz05}.  These overdensities collapse to ultimately form stars, and along the
way form \emph{prestellar cores}.  The purpose of this series of papers is to
examine the formation of these prestellar cores, and in this initial installment
we focus on the initial conditions of the cores;  What are the properties of the gas that becomes a prestellar core?

Prestellar cores are cold ($T=10-20$K), dense ($n\sim 10^4\percc$) small
($r=0.1$pc) somewhat round objects that are often associated with young stellar objects and
star forming regions \citep{Guzman15}.  They form the basic reservoirs out of which stars can
form, and thus their formation process will tell us a great deal about the
formation of stars.  Within the prestellar core, a protostar is formed along
with a disk and 
possibly a jet that will further influence the final properties of the star
\citep{Matzner99}, but
this late collapse is beyond the scope of the current work.  Here we focus on
the portion of the collapsing molecular cloud that ultimately ends in the
prestellar core.

The collapse of molecular clouds is a messy process, and many aspects are under
investigation.  In broad strokes, some portion of the cloud becomes unstable to
gravity, as forces that are not gravity give way to forces that are gravity.
That portion then can gain mass from the surrounding cloud as it collapses.
The final mass of a prestellar core is the sum of the mass of the initial fragment 
plus any gas it gains during the collapse, minus any gas it loses due to
turbulence and tidal interactions with other bits of the cloud.
In the \emph{turbulent fragmentation} model, the mass and collapse rate is largely
determined by the initially unstable portion of the cloud \citep{Krumholz05,
Hopkins13}.  In the \emph{competitive
accretion} model, the mass is largely determined by the later tidal
interactions \citep{Bonnell01}.  In a new model, the \emph{inertial flow} model, the mass is
determined by the large scale velocity patterns in the cloud \citep{Padoan20,Pelkonen21}.  Finally, the
\emph{stochastic star formation} models treat the collapse of clouds as a random
walk  \citep{Scannapieco18}.   It is likely that none of these are individually correct, but as we will
show the collapse has aspects of all of these processes.

The discussion of the formation of stars is filled with discussions of fractals
and filaments \citep{Elmegreen96,Hacar22}.  Clouds are observed to have fractional dimensions
of 2.6-2.8 \citep{Stutzki98} or are perhaps better described by a multifractal
system \citep{Yahia21}.  Elongated filamentary structures within a cloud will have fractal
dimensions less than that.  Ultimately, the formation of stars is one of
dimension reduction;  the three-dimensional cloud must be forced into a
zero-dimensional star by way of $~2$ dimensional shocks and $~1$ dimensional
filaments.

In this work, we simulate collapsing molecular clouds by first stirring a
periodic box with a large-scale, supersonic driving pattern.  
Once a statistically steady state is reached, massless tracer particles are
inserted uniformly throughout the box, and gravity is turned on.
  This
onset of gravity is $t=0$ for our purposes.  The
cloud is then allowed to collapse.  Particles in dense regions at the end of the
simulations are identified, and traced back to the initial phase at $t=0$.
Exact details of the simulation setup, particle advection, and particle
selection can be found in Section \ref{sec.sims}.

Other works have also examined the collapse of a cloud in a
Lagrangian manner.  \citet{Mocz18} examined the behavior of magnetic fields
during collapse.  \citet{Kuznetsova19} examined the angular momentum of
collapsing objects.  \citet{Pelkonen21} also used tracer particles to examine
accretion onto cores.   \citet{Smullen20}  used consecutive isocontours of
density to follow cores, and showed that isocontours are not structures that can
be used for consistent identification of cores.

This is the first paper in a series.  In the second paper, we will examine the
rate and geometry of the collapse.  In the third, we will focus on the magnetic
field behavior.

This paper is organized in the following manner.  We present the code and
simulations in Section \ref{sec.methods}.  We present results in Section
\ref{sec.results}; Section \ref{sec.demographics} shows projections and
path lines for the cores; Section \ref{sec.pdfs} presents the probability
distribution function (PDF) for density, speed, magnetic field, and
gravitational potential; Section \ref{sec.means} presents the second order
structure function;   Section \ref{sec.length} presents the length scales of the
simulation; Section \ref{sec.hulls} presents the convex hull bounding of each
preimage and discusses their contents; Section \ref{sec.fractals} presents the
fractal dimension of the preimages. We present a 
new model for the star formation rate in Section \ref{sec.discussion}, and
summarize
in Section \ref{sec.conclusions}.

\begin{figure*} \begin{center}
\includegraphics[width=\textwidth]{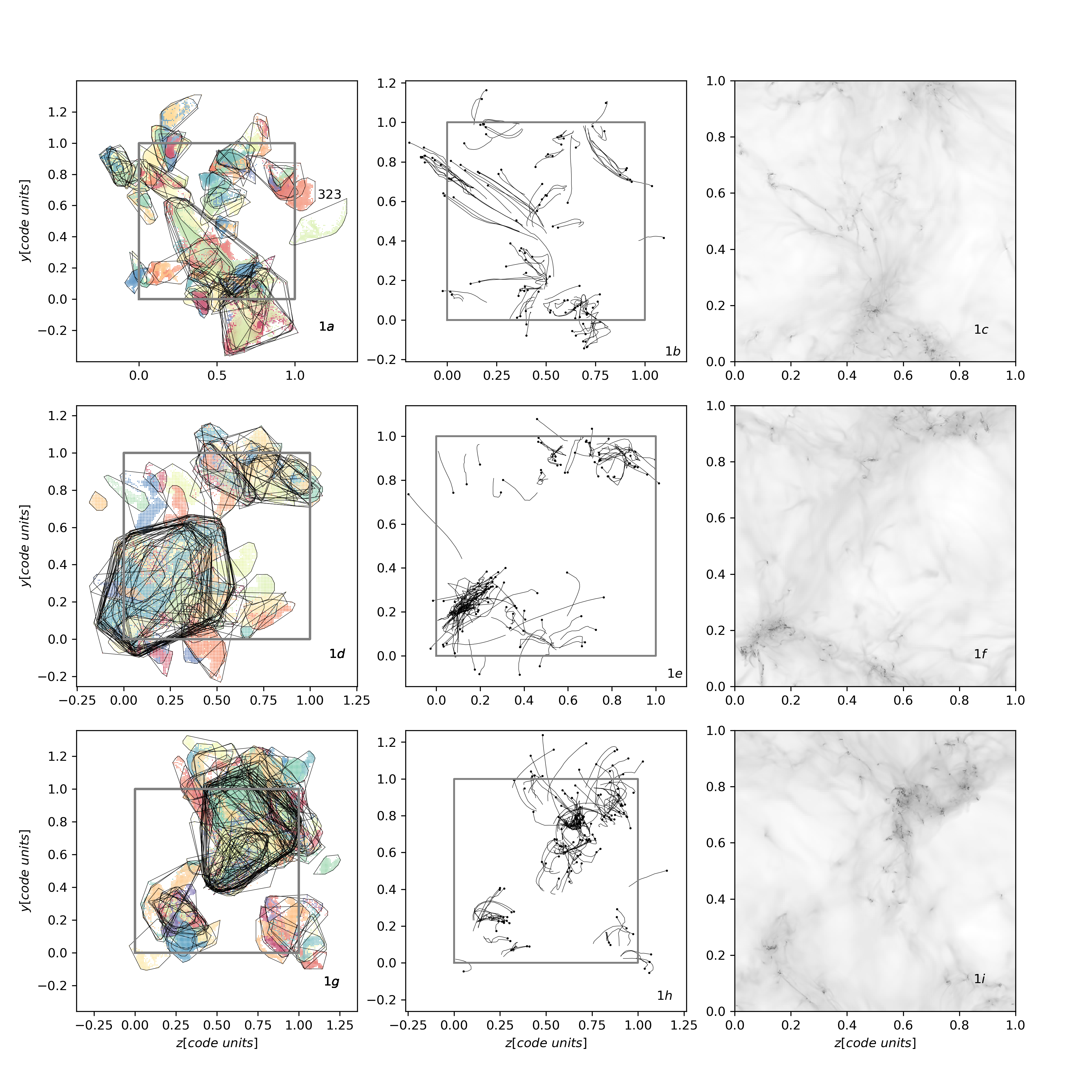}
\caption[ ]{The evolution of our simulation. Simulation $\sima$ is in the top row, $\simb$ in the middle row, and $\simc$ in the bottom row.
\emph{(Left Column)} Preimages of the cores.
Colored pixels represent particles that will end up in dense zones.  Most of the overlap between different
preimages is not due to projection effects, but show preimages that occupy
similar space.  \emph{(Middle Column)} Tracks of the centroids for each core as
they collapse through time,
mapping the first column to the third.  Each track starts at the black point. \emph{(Right Column)} Projections of the final frame with hundreds of dense knots. }
\label{fig.proj} \end{center} \end{figure*}

\begin{figure*} \begin{center}
\includegraphics[width=0.24\textwidth]{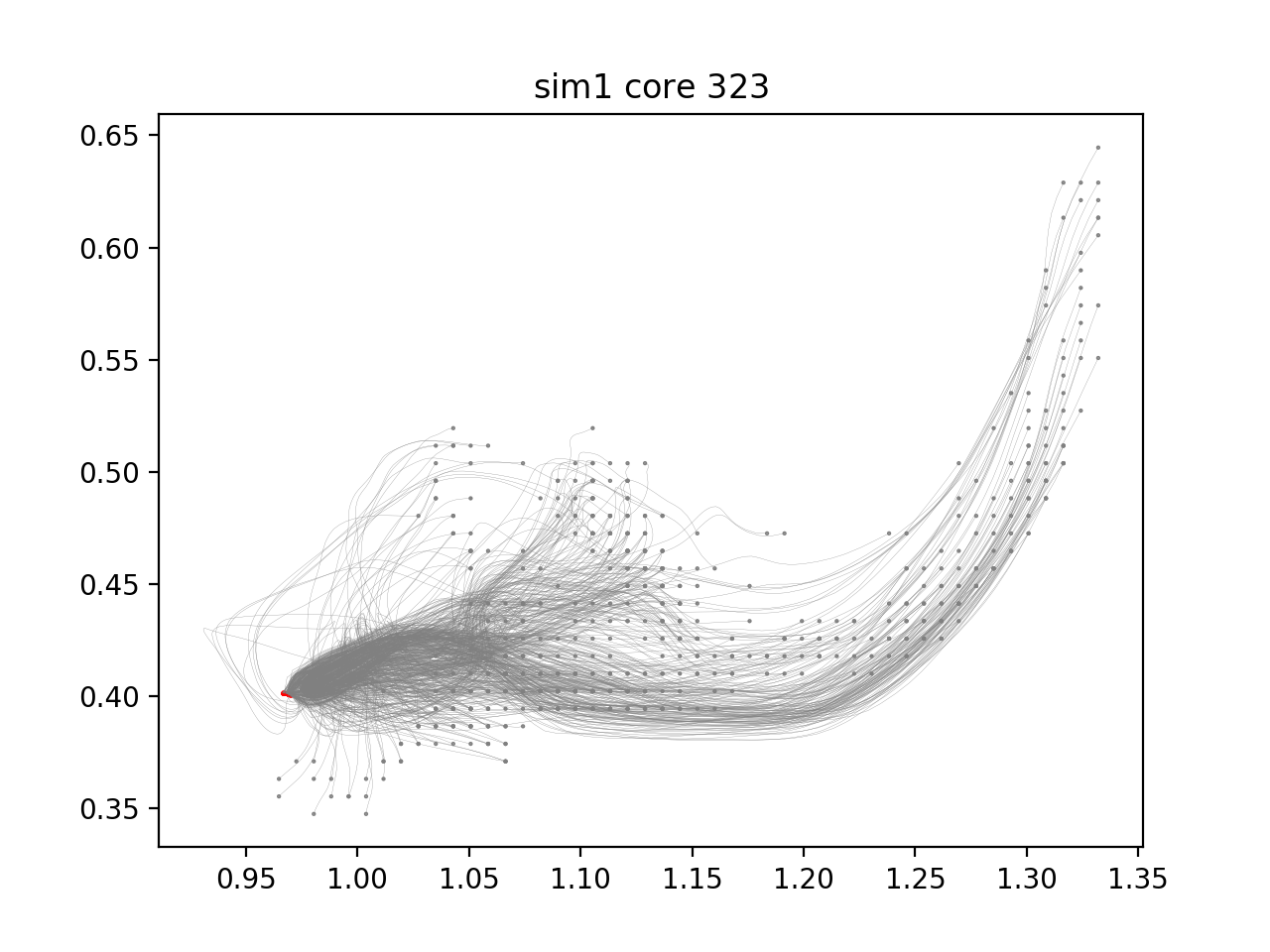}
\includegraphics[width=0.24\textwidth]{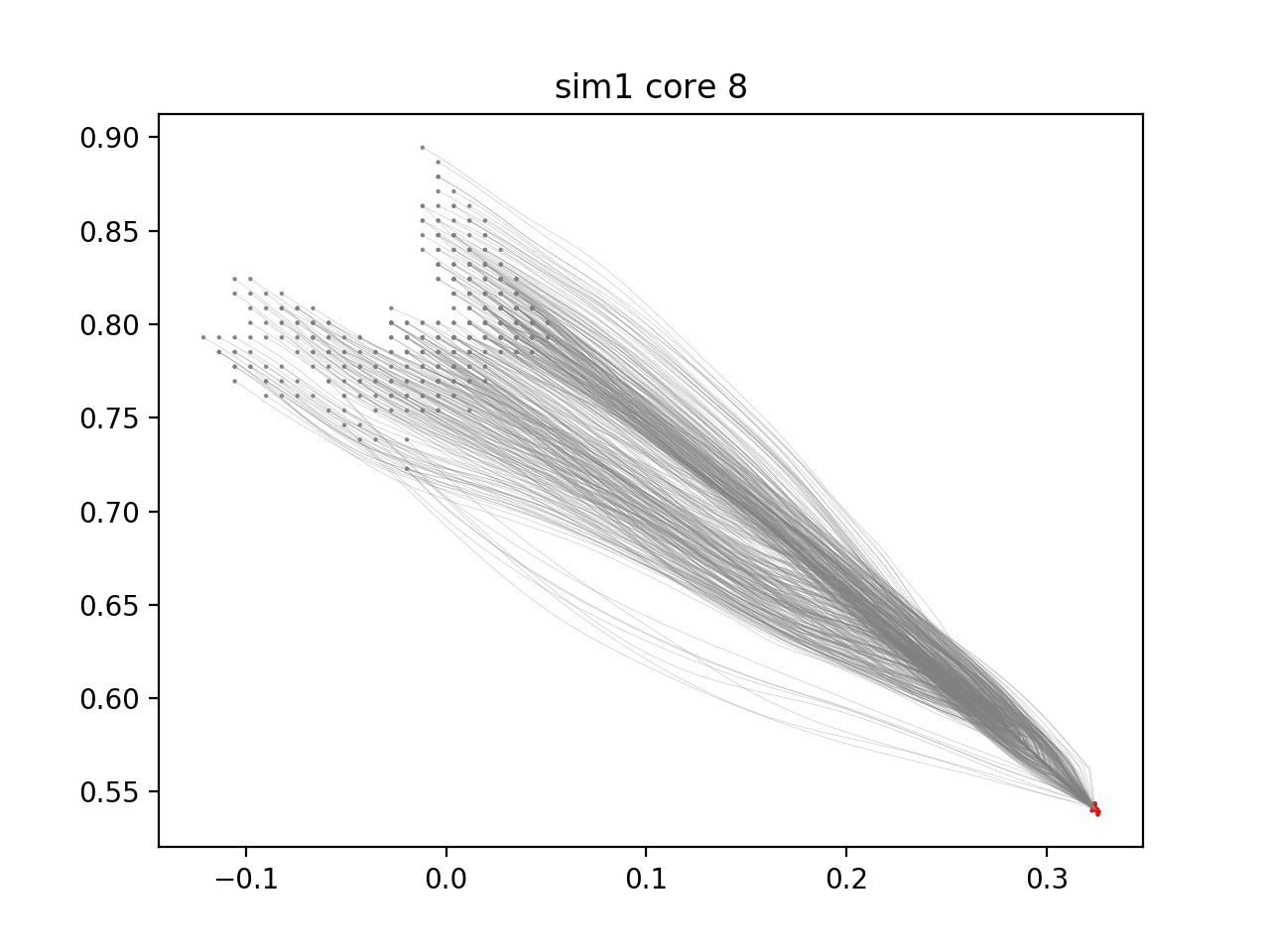}
\includegraphics[width=0.24\textwidth]{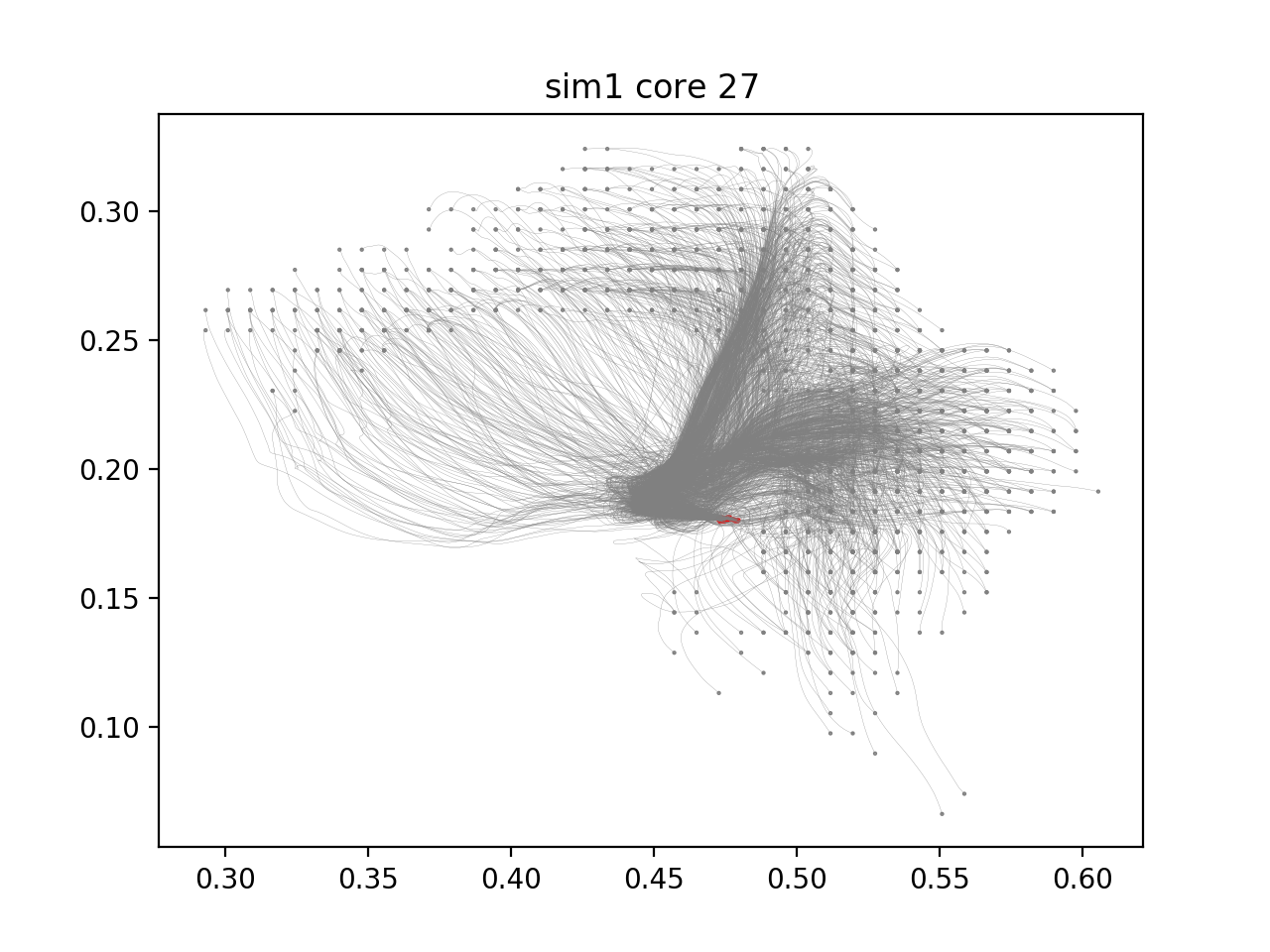}
\includegraphics[width=0.24\textwidth]{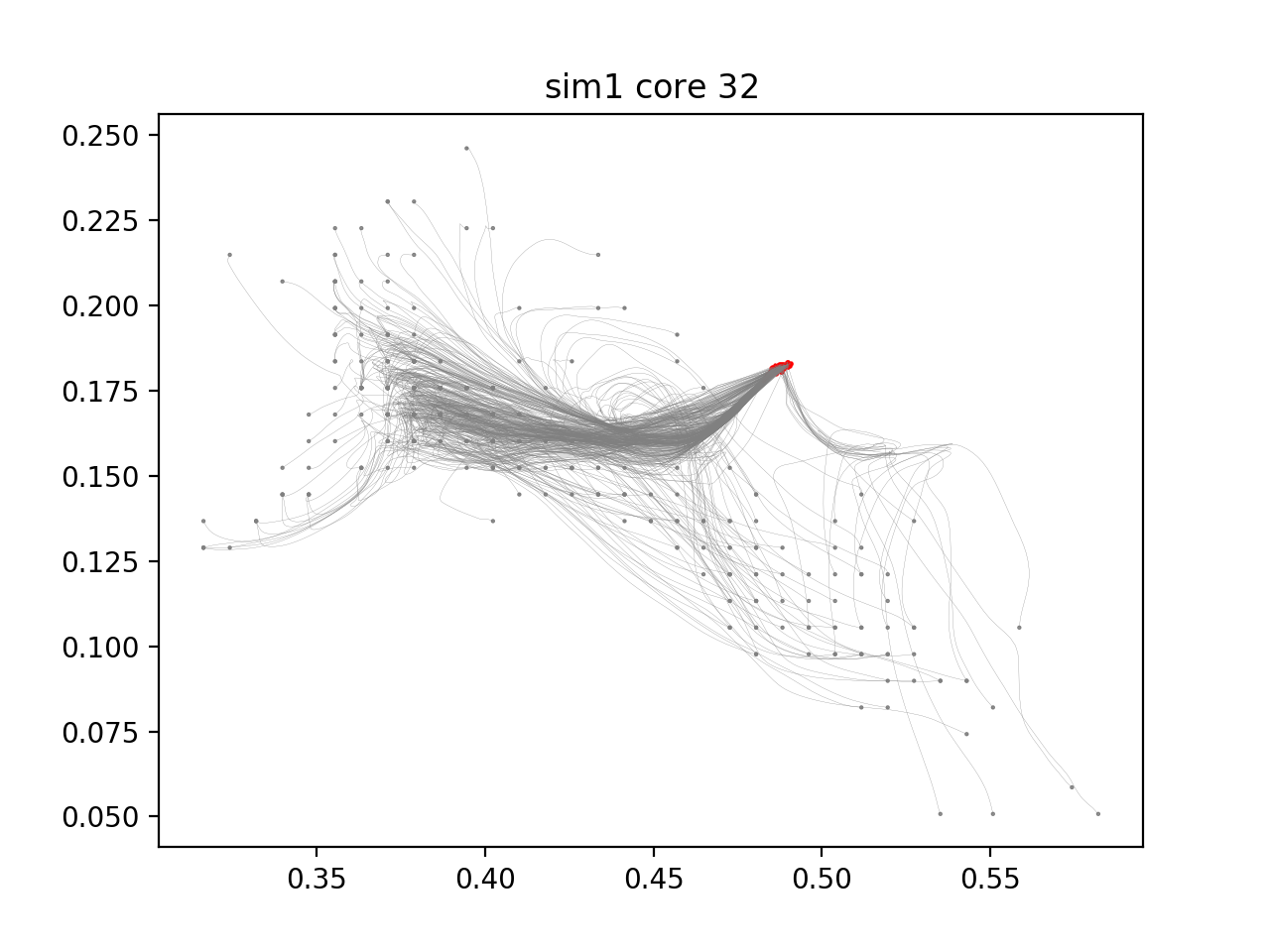}
\includegraphics[width=0.24\textwidth]{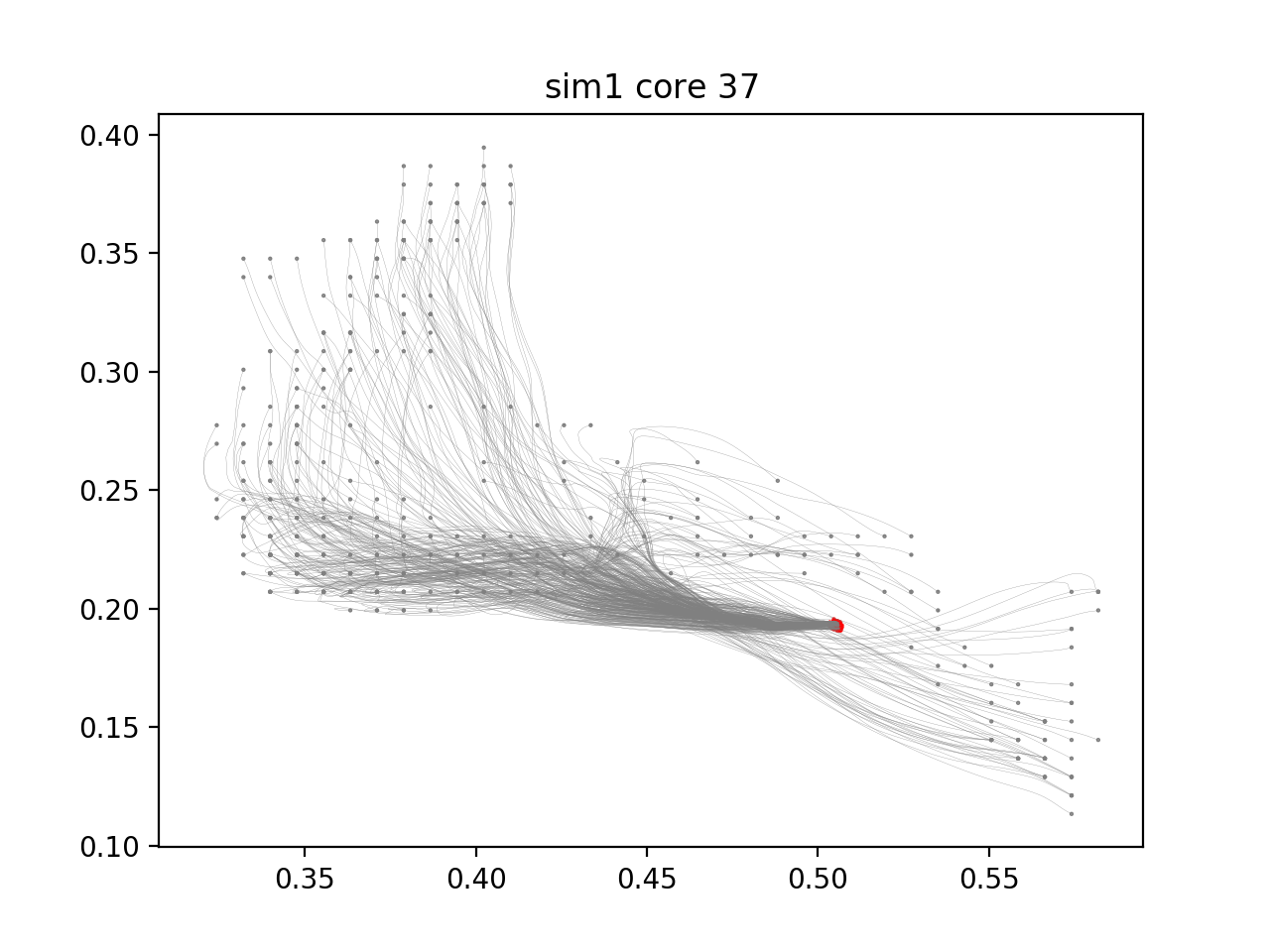}
\includegraphics[width=0.24\textwidth]{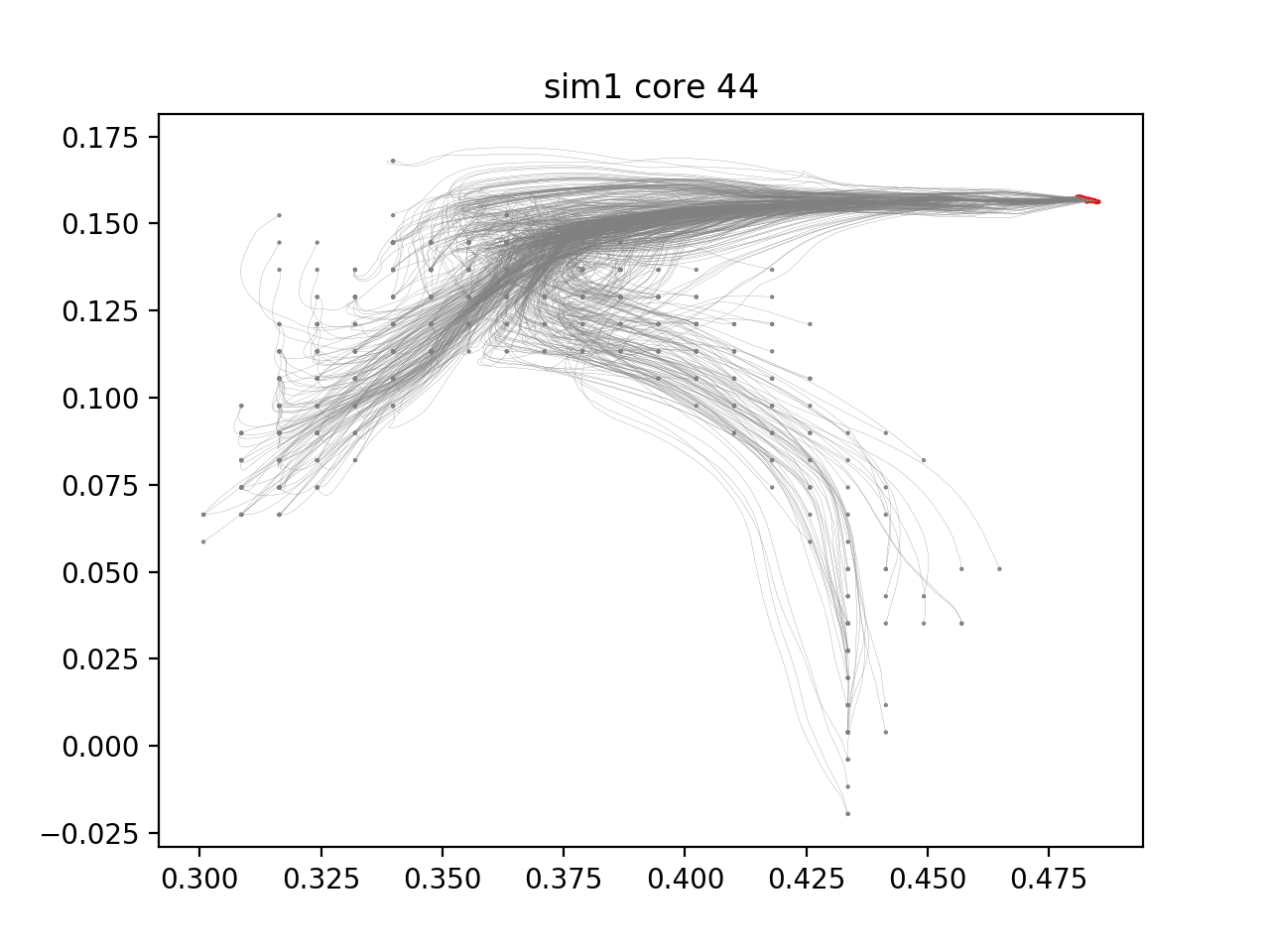}
\includegraphics[width=0.24\textwidth]{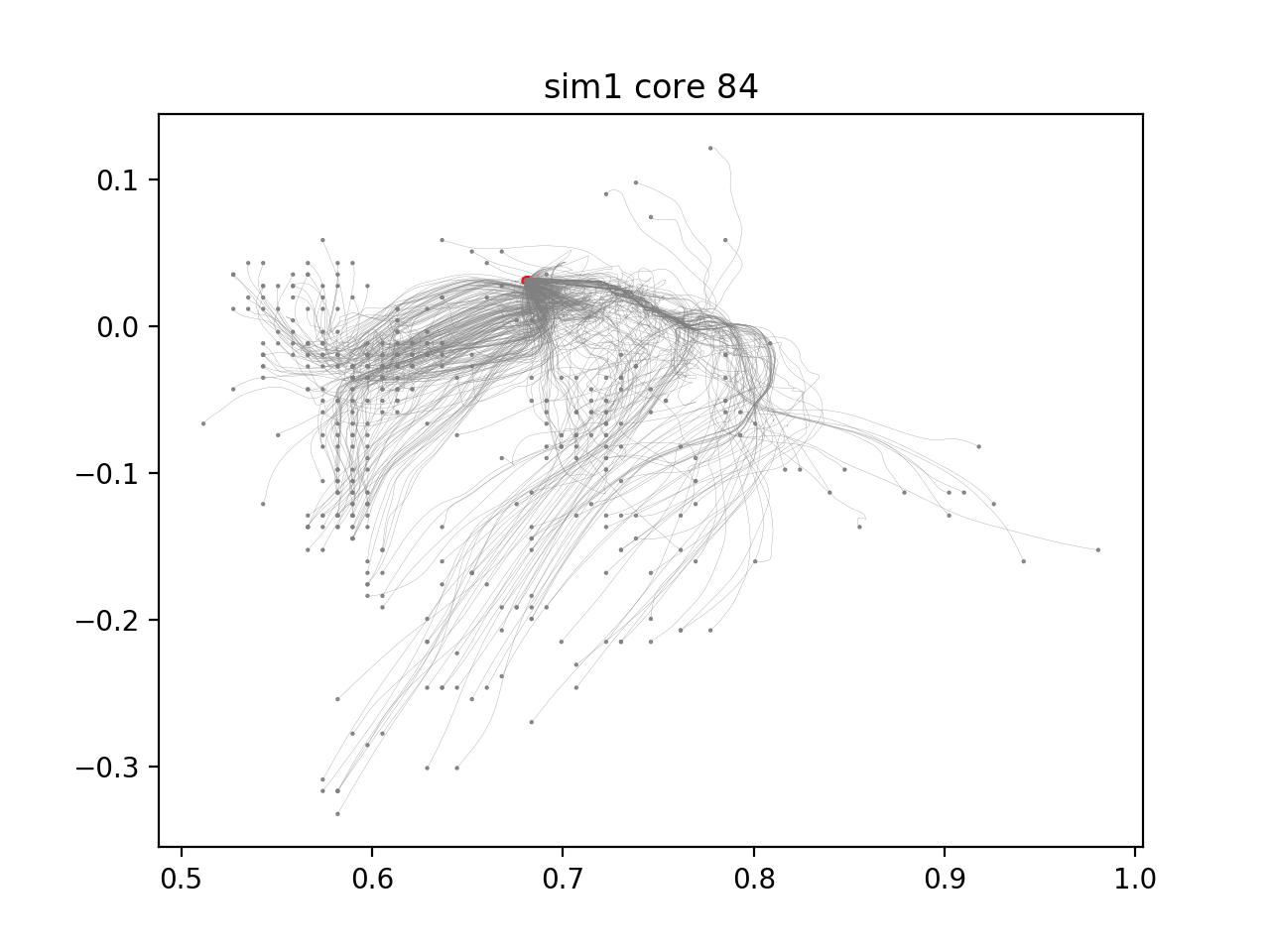}
\includegraphics[width=0.24\textwidth]{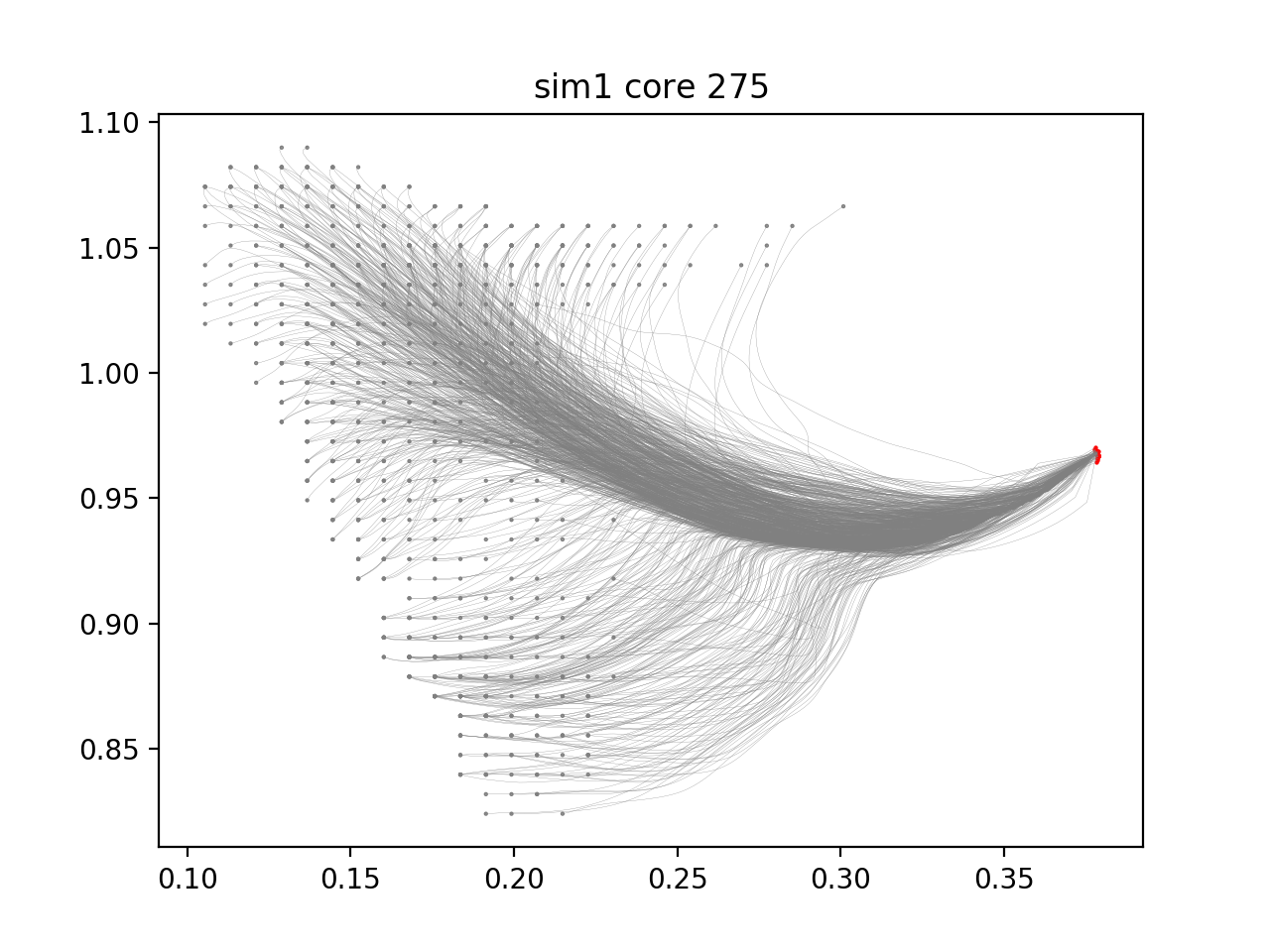}
\includegraphics[width=0.24\textwidth]{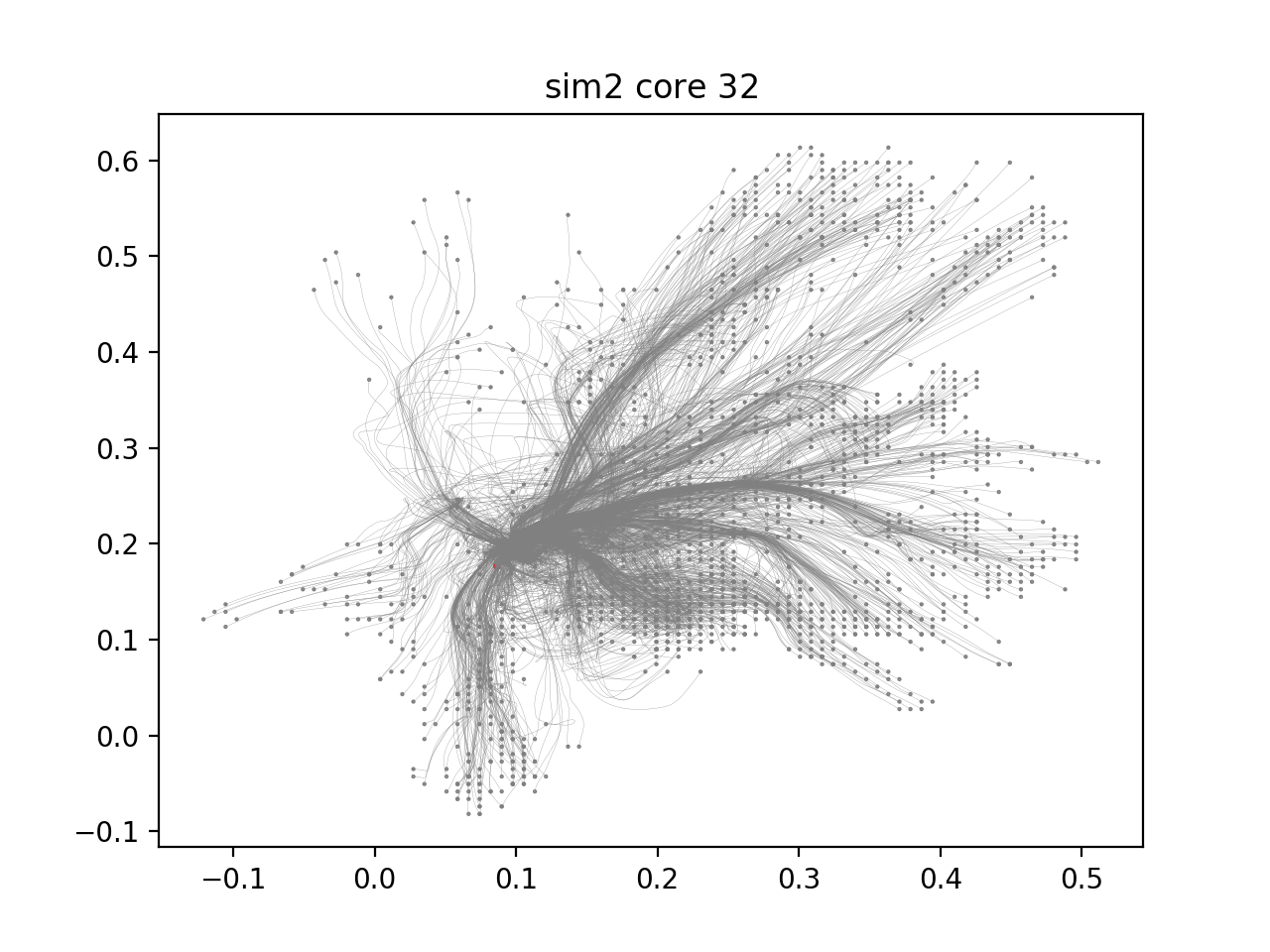}
\includegraphics[width=0.24\textwidth]{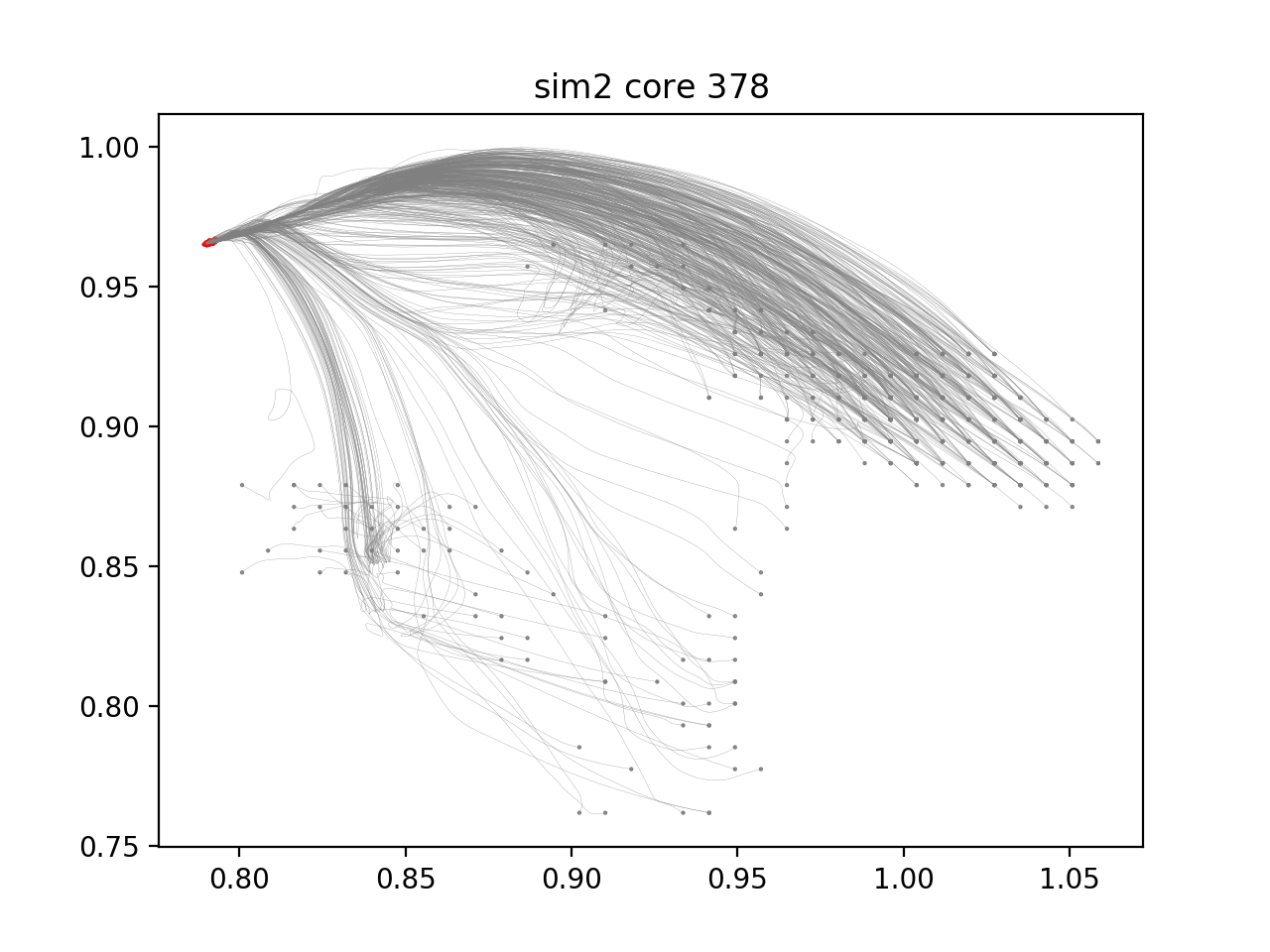}
\includegraphics[width=0.24\textwidth]{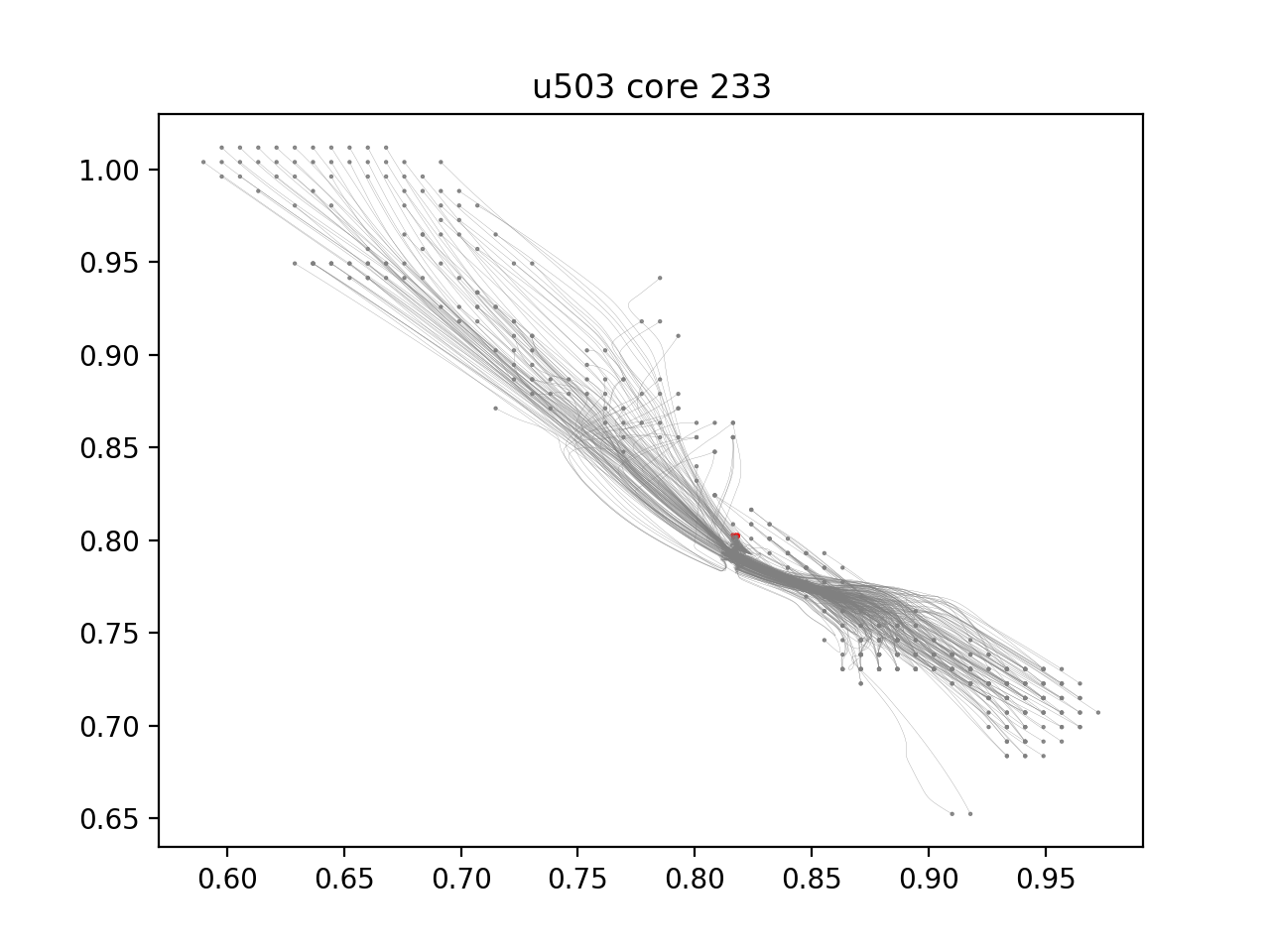}
\includegraphics[width=0.24\textwidth]{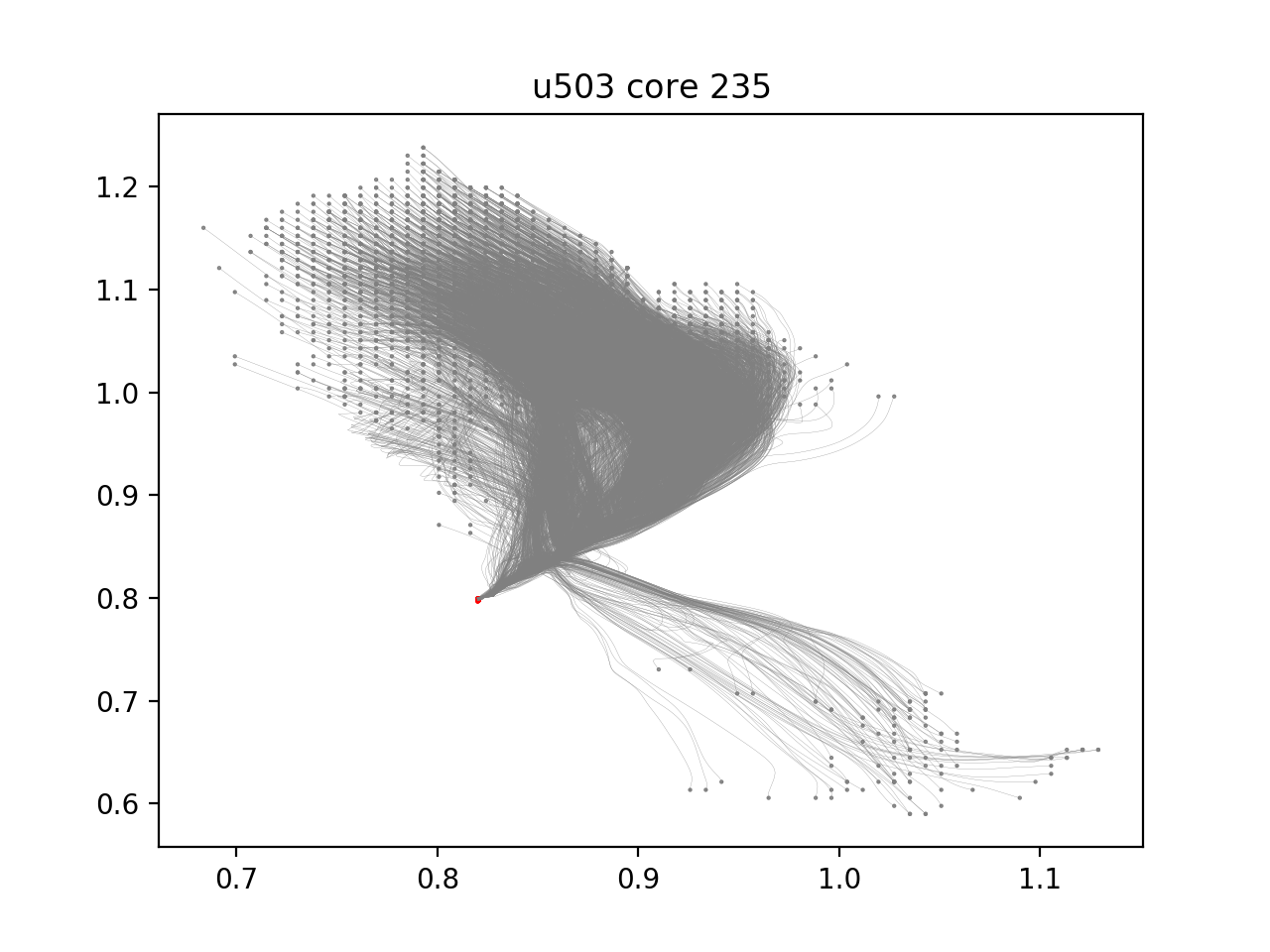}
\caption[ ]{Particle tracks for various cores from $t=0$ (black dots) to
$t=t_{\rm{f}}$ (red dots).  Cores exhibit a variety of behaviors and distributions
as they collapse. Some tracks are rather straight, and signify direct collapse
with perhaps a bulk motion. Many show more complex dynamics.  Axis labels have
been omitted for clarity, in all plots the horizontal axis is the $y$ position in code
units, and the vertical axis is the $z$ position. It should be noted that the scale of each core is
different; some take up half the domain (core 84, core 32) but others are much
smaller (core 44).}
\label{fig.hair} \end{center} \end{figure*}

\section{Methods}
\label{sec.methods}

We run three simulations of adaptive mesh refinement (AMR) magnetohydrodynamic 
(MHD) simulations using the code Enzo \citep{Collins10, Bryan14}.   
The simulations use $128^3$ root grids and 4 levels of refinement, and $128^3$
tracer particles.  We employ isothermal magnetohydrodynamics and gravity.
Here we will describe the code and simulations.

\subsection{Enzo}
\label{sec.code}

Enzo \citep{Collins10, Bryan14} is an astrophysical hydrodynamics code that
uses the AMR strategy of \citet{Berger89} to dynamically add patches of additional
refinement as needed by the simulation.  It has been used for hundreds of
astrophysical publications, including star formation \citep{Collins12} and
reionization and structure formation \citep{Xu16}.

In this work, we use ideal isothermal MHD, AMR, and self gravity.  We use the
piecewise linear method of \citet{Li08a}, the isothermal Riemann solver from
\citet{Mignone07}, the constrained transport method of \citet{Gardiner05}, and the
divergence-preserving AMR algorithm of \citet{Balsara01}.  Self gravity is
handled with a fast Fourier transform (FFT) method on the root grid, and a multi-grid method on the
subgrids.  Tracer particles integrate their position by first using a trilinear
interpolation of the velocity field, and a drift-kick-drift method of advance.
More details on the tracer particles can be found in Section \ref{sec.tracers} and
\citet{Bryan14}.
More details on the solvers can be found in \citet{Collins10} and \citet{Bryan14}.

\subsection{MHD simulations}
\label{sec.sims}

Our simulations began by driving periodic boxes with large scale (wavenumbers
between 1 and 2) solenoidal velocity patterns, with a target sonic Mach number
$\mathcal{M_{\rm{s}}}=v_{\rm{rms}}/c_s = 9$, where $v_{\rm{rms}}$ is the root mean square
velocity, and $c_s$ is the sound speed.  The driving was continued for several
dynamical times ($t_{\rm{dyn}}=L_0/2 v_{\rm{rms}}$, where $L_0$ is the box size)
until a statistically steady state was achieved.  This driving was done with a
box of $1024^3$ zones per side.  

The box is then down-sampled to $128^3$, tracer particles are added (one
particle for each zone) and the simulation is restarted with
gravity and 4 levels of AMR.  For these simulations, refinement is triggered whenever the local
Jeans length is less than 16 zones (i.e. $L_{\rm{J}} = \sqrt{ \pi c_s^2 /G
\rho} < 16 \Delta x$). 
Collapse continues for about 0.8 free-fall time ($\tff=\sqrt{3 \pi/32 G \rho_0}$), where $\rho_0$ is the average density.
A summary of the simulations can be found in Table \ref{tab.table1}

\begin{table}
\begin{center}
\caption{Simulation parameters and number of cores.  All simulations have a mean
Mach number of 9 and mean virial parameter $\alpha=5 \sigma^2 R / G M=1$.
Plasma beta, the ratio of thermal to magnetic pressure, $\beta =
P_{\rm{gas}}/P_{\rm{mag}}$, and number of cores found $N_{\rm{cores}}$, are
below}
\begin{tabular}{l l l}
    \label{tab.table1}
Name & $\beta$ & $N_{\rm{cores}}$\\
\hline
\sima & 0.2 & 113\\
\simb & 2.0 & 112\\
\simc & 20  & 136
\end{tabular}

\end{center}
\end{table}

Unlike many star formation simulations in the literature, we do not employ sink
particles in this work.  This choice was made as to not influence the
small-scale dynamics of the collapse with additional model parameters.  
This limits the run time of our simulations, as the code is not able
to handle real singularities.

These simulations are scale free, so most analysis is done relative to the mean
quantities in the box (e.g. mean density and sound speed).  However it is useful
to associate the results with physical units.  We follow the scaling laid out in
\citet{Collins12}; the length scale of the box is 4.6 pc, the density is 1000
$\percc$, the time scale is 1.1 Myr, the mass of the box is 5900 $\msun$, and
the magnetic field strength is (13, 4.4, 1.3) $\muG$ for (\sima, \simb, and
\simc).  

\subsection{Tracer Particles}
\label{sec.tracers}

\def\ex{\ensuremath{\epsilon_x}}
\def\ey{\ensuremath{\epsilon_y}}
\def\ez{\ensuremath{\epsilon_z}}
Our tracer particles follow the flow by way of a tri-linear
interpolation of the velocity field on the grid.  Thus if a particle is between
the mid-points of zones $(i,j,k)$ and $(i+1,j+1,k+1)$, its velocity in each
direction is a linear combination of the velocities in the 8 neighboring zones.
If the particle's position is $(x,y,z)$ and the zone center is at
$(x_i,y_i,z_i)$, we define the distance to the left point
as $\ex=(x-x_i)/\Delta x$, $\ey=(y-y_i)/\Delta y$,
$\ez=(z-z_i)/\Delta z$, and the distance to the right point is $1-\ex$.  The
full interpolation is then
\begin{align}
v_x(x,y,z) 
 &= v_{x,i,j,k} \ex \ey \ez+\\
 &= v_{x,i,j+1,k}(\ex)(1-\ey)(\ez)+\nn\\
 &= v_{x,i,j,k+1}(\ex)(\ey)(1-\ez)+\nn\\
 &= v_{x,i,j+1,k+1}(\ex)(1-\ey)(1-\ez)+\nn\\
 &= v_{x,i+1,j,k}(1-\ex)(\ey)(\ez)+\nn\\
 &= v_{x,i+1,j+1,k}(1-\ex)(1-\ey)(\ez)+\nn\\
 &= v_{x,i+1,j,k+1}(1-\ex)(\ey)(1-\ez)+\nn\\
 &= v_{x,i+1,j+1,k+1}(1-\ex)(1-\ey)(1-\ez)\nn.
\end{align}
Here $v_{x,i,j,k}$ denostes the numerical value of the x-velocity in the zone
$(i,j,k)$, and $v_x(x,y,z)$ is the continuous particle velocity.  Similar
expressions hold for $v_y(x,y,z)$ and $v_z(x,y,z)$.  

The particle position is updated from this in a kick-drift manner,
\begin{align}
x^{n+1} = x^{n} + \Delta t v_x.
\end{align}

One of the advantages of this
algorithm is that it is low order, so it injects no complexity into the flow on its own. Particles
that are closer than $\Delta x$ have velocity increments that are linear in the
spacing. One consequence of
this is the over clustering of particles \citep{Konstandin12}.  Given two particles with separation less
than $\Delta x$,  if the flow is converging, the particles will be driven to a
separation of zero in a matter of a few time steps.  This is easiest to see in 1d,
where
\begin{align}
v_{x,1d} = v_i \ex + v_{i+1} (1-\ex).
\end{align}
Two particles with separations $D \Delta x$ will have a relative velocity, $v_r$, of
\begin{align}
v_r = (v_{i+1}-v_i) D,
\end{align}
and will have zero spacing in a time 
\begin{align}
t=\frac{D \Delta x}{v_r}=\frac{1}{\nabla \cdot v}.
\end{align}
Because of this, the density of the particles is not a reasonable representation
of the density of the gas, and we do not consider it.  We only use the particles
to identify which zones to analyze, and we perform all analysis on the Eulerian
zone data.  We discuss particle selection in the next section.

\subsection{Particle Analysis}
\subsubsection{In Brief}

In order to study the gas that \emph{will} collapse,
 we must first identify
the gas that \emph{did} collapse.  That is, we need to identify the dense cores,
$c_i$, at the
end of the simulation, and select the particles within that core to examine at
earlier times.  We discuss peak finding and particle selection in Section
\ref{sec.identify}.  Once particles for $c_i$ have been selected at
$t=t_{\rm{final}}$, their location
in earlier snapshots can be found.  The zones that contain particles in core
$c_i$ at $t=0$ is referred to as the \emph{preimage} gas.

In all analysis presented here, the tracers only identify the
zones to analyze; all physical quantities are taken from the grid data, not the
properties of the particles.  Further, each zone is only counted once; if two
particles reside in the same zone, that zone is counted only once for averaging
purposes.  

One of the guiding principles of this study is to avoid defining the ``edge'' of
a core.  There are a number of possible ways that the ``edge'' of a core has
been defined in the literature,
e.g. density isocontours with suitable energy contents.
However, lacking a predictive theory
of turbulence, it is impossible to say a priori if the gas within the ``edge'' of the
core
actually ends on the star.  In fact, we argue that any closed 2d surface will
necessarily contain a substantial fraction of gas that does \emph{not} end on
the star.  

\subsubsection{Particle selection}
\label{sec.identify}

We first identify density peaks in the last frame of each simulation using {\tt yt}.
This collection of density peaks contains all of the cores we are
interested in, as well as a number that are low-density turbulent fluctuations.
Peaks formed by gravity are,
in these simulations, easily distinguishable from those formed by the
turbulence, as they have typical densities of $\rho_{peak}\sim 10^{5...7}$, while the
turbulent peaks have $\rho_{peak}\sim 10^{2...3}$.
The distribution of peak densities was found to be bimodal, with one population
having $\rho_{peak} << 10^4 \rho_0$, and one population with $\rho_{peak} >>
10^4\rho_0$.   We retain only peaks
in the upper portion of the distribution.   Turbulence at this Mach number is
incapable of making densities above a few hundred, anything truly dense can be
only formed by gravity. Not all of our cores will survive infancy, but this is
not the concern of the present work.  

Once the peak zones are identified in space, the particles around them can be
selected.
Owing to the fact that particles
cluster more tightly than the gas, the particles
reside only in the densest zones around the peak.  It is found that
particles all reside well within a density contour of
$\rho_{select}=\rho_{peak}^{3/4}$, so we use this to select particles for each
density peak.  A density contour at $\rho_{select}$ is used to identify particles to track.  The density contour is otherwise ignored.

It should be emphasized that particles are not selected by taking density
contours.  Any isosurface of density is absolutely meaningless as far as a
turbulent cloud is concerned.  Our particles are selected by virtue of being
located at density maxima.  

As a test of the particle selection technique,  all of the analysis presented here was performed with a more
restrictive particle selection.  This more restrictive selection takes only
particles in the zone containing the densest peak and its immediate neighbors.
This gave fewer particles, but did not change any of the results presented here.

\subsubsection{Analysis}

Once particles for each core $c_i$ are identified, their locations in previous
frames are identified. The collection of zones marked by tracers for a
core $c_i$ is referred to as the \emph{preimage} $P_i$.  

It has been seen by us and others that particles, lacking pressure, cluster
tighter than the gas does. 
Thus we do not at any
point analyze the density
or velocities of the particles themselves; particles
are only used to identify grid zones to analyze.  
Unless otherwise noted, we only
count each zone once, even though there are eventually multiple particles in
each zone.  This does not affect the results in this work, as we are
primarily focusing on the first snapshot where there is an exact 1-1
correspondence between zones and particles.  As we will discuss in \papertwo,
the number of unique zones does not decrease significantly until the last few frames.

\subsubsection{Online Core Browser}

We form hundreds of cores, but a publication is finite.  In order to enjoy the
rich spectrum of cores that are formed along their collapse trajectories, we have
built an online database of cores and their properties.  This can be found at
\url{http://cores.dccollins.org}.

\section{Results}
\label{sec.results}

\subsection{Core Demographics}
\label{sec.demographics}

\begin{figure*} \begin{center}
	\includegraphics[width=0.99\textwidth]{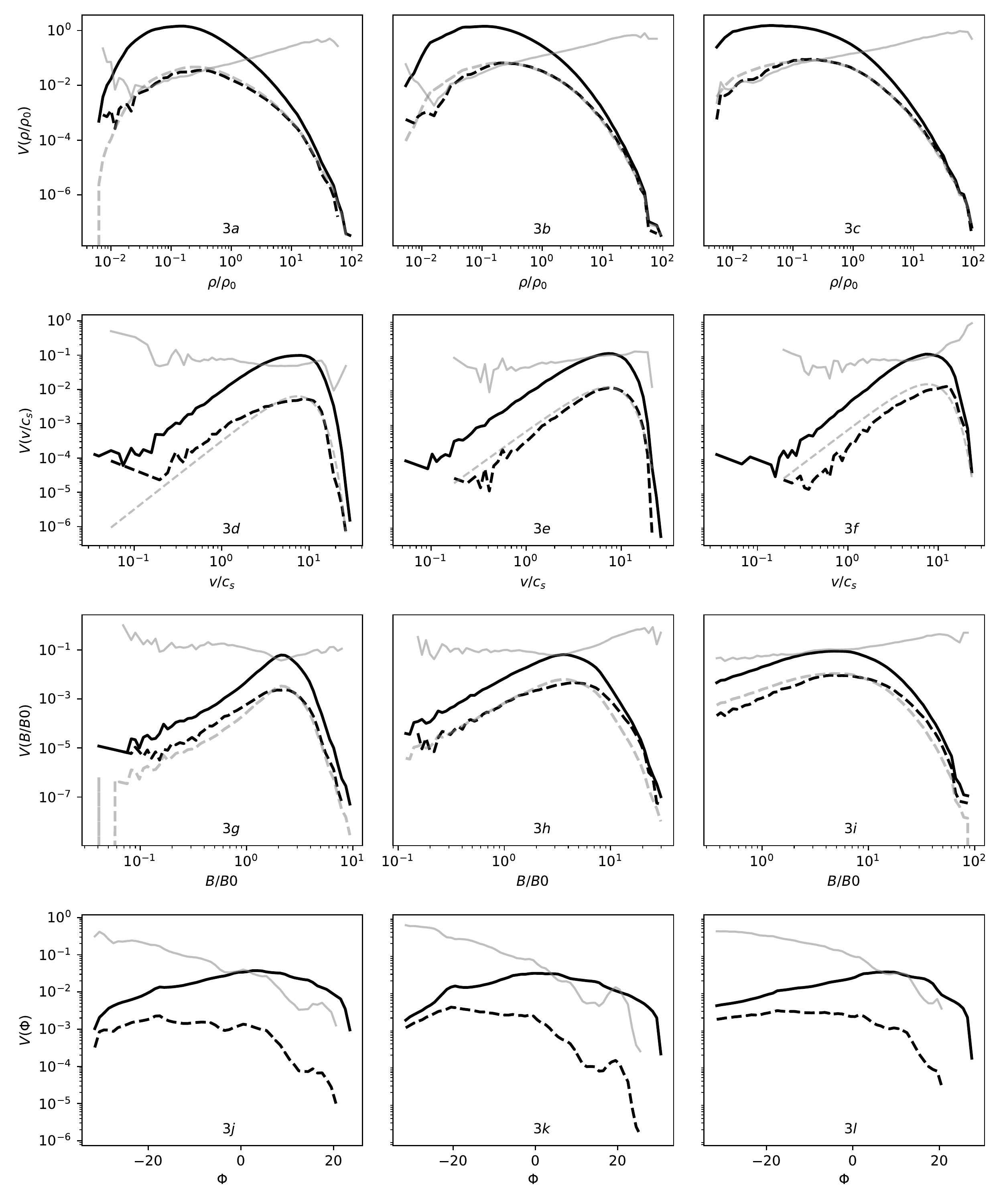}
\caption[ ]{The probability density functions (PDF) for $\rho$ (top row, 3a-3c) speed $v$ (second row, 3d-3f), magnetic field strength $B$ (third row, 3g-3i)
and gravitational potential $\phi$ (bottom row, 3j-3l).  Each panel shows 4
lines, each at time $t=0$. The solid black line shows the PDF for all gas,
$V(q)$. The
dashed black line is the distribution for all preimage gas, $V(q|*)V(*)$.  The
solid grey line shows the probability of forming a core at a given value,
$V(*|q)$.  The dashed grey line is the prescription for the
preimage 
distribution, where available (see text for details).   This has not yet been found for the potential field.  Simulation
 \sima\ is shown in the first column, \simb\ in the second, and \simc\ in the
 third column. }
\label{fig.pdfs} \end{center} \end{figure*}

\begin{figure} \begin{center}
\includegraphics[width=\hw\textwidth]{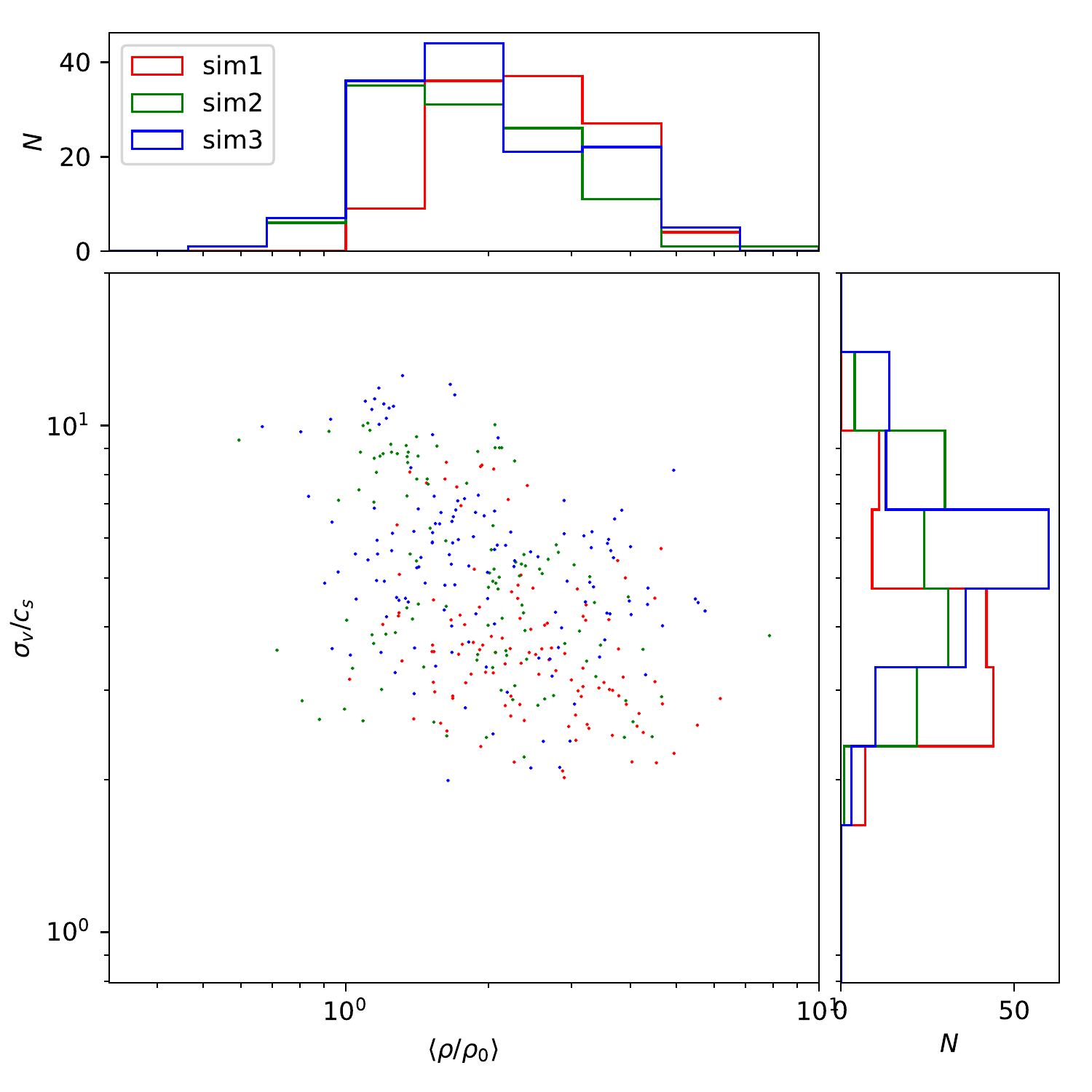}
\caption[ ]{The mean density $\langle \rho/\rho_0\rangle$ and rms velocity
$\sigma_v$ for all preimages for each simulation.  No correlation is seen between
the two.  The density includes cores whose initial mean density is below the
mean of the box.  All cores are supersonic. }
\label{fig.means} \end{center} \end{figure}

Figure \ref{fig.proj} shows a summary of our results.  
The top, middle, and bottom rows show \sima\, \simb\, and \simc, respectively,
which have initial plasma $\beta=0.2, 2.0, 20.0$.  The three columns show the
beginning of the simulation on the left, the end of the simulation on the right,
and the path between the two in the center.  We will elaborate on each of these.

The first column (\ref{fig.proj}a, \ref{fig.proj}d, and \ref{fig.proj}g)  shows
the initial positions of the preimage particles.  Each colored cloud of points
forms a distinct separate core at the end of the simulation.  
The grey square denotes the boundary of the
box, and particles have had periodic jumps "straightened out" so they appear to
come from outside the box.  The contours surrounding each preimage denotes the
\emph{convex hull}, which will be discussed in more detail in Section \ref{sec.hulls}.
The substantial overlap between different preimage clouds is not the result of
projection, but real overlap.  This will be discussed in Section
\ref{sec.overlap}.

The middle column (\ref{fig.proj}b, \ref{fig.proj}e, and \ref{fig.proj}h) shows the tracks
of the centroid of the core as it collapses, the dot marks the beginning of the
track.  A variety of behaviors can be seen.  Some are long and relatively
isolated, while some are tangled together in clusters.  The wide variety of collapse
morphologies can be seen in Figure \ref{fig.hair}, and will be revisited in
\papertwo.

The third column (\ref{fig.proj}c, \ref{fig.proj}f, \ref{fig.proj}i) shows projections of the entire box at the end of
each simulation.  
Dense knots are clearly visible as black spots.  We identify (323, 381,
295) density peaks in (\sima, \simb, and \simc), respectively, of which (113,
112, 136) survive our density cuts.

Insight about the collapse process can be had by plotting  particle
trajectories for individual cores, as seen in Figure \ref{fig.hair}.  This shows a collection of
pathlines for every particle for several cores.  The particles begin at the black
dots (which are regular by construction) and end on the red dots.  There is
quite an array of collapse modalities seen by the collapsing cores.  
Core 323 (top left corner of Figure \ref{fig.hair}  as well as \ref{fig.proj}a) seems to
be a filamentary structure, and the flow starts in the direction of the
filament; core 8 moves coherently while collapsing.  Core 27 comes from a
complex region, and has a more complex accretion pattern.  
In the bottom row, \simc\ core 32 seems to form from multiple sub-clumps, while
core 378 has two major sources of gas that merge at the end.  
Similar plots can be found for every core in our online core browser.

\subsection{PDFs of Preimage Gas}
\label{sec.pdfs}

\begin{figure*} \begin{center}
\includegraphics[width=\textwidth]{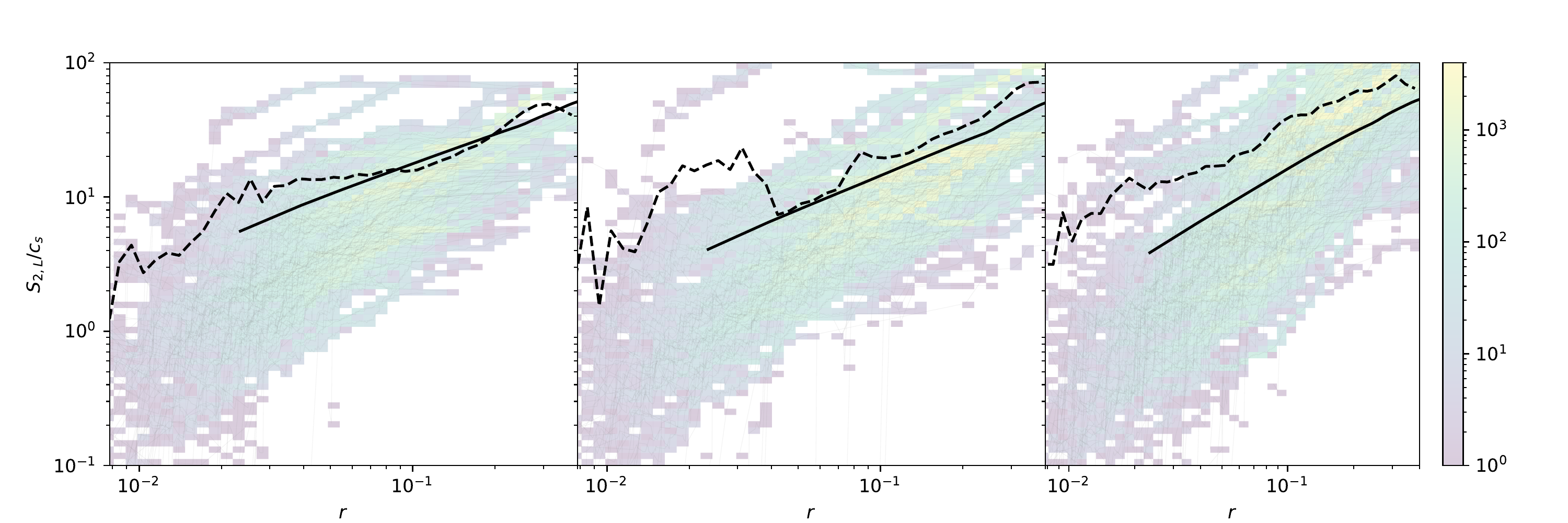}
\caption[ ]{The second order velocity structure function.  The thick
black line is \sss\ for the entire simulation (the typical definition).  The
faint thin black lines show \sigmavl\ for each core, $c_i$.  The
color map is a heat map, a histogram of each track that goes through each point.
The dashed black line shows the mean of \sigmavl\ over all cores.}
\label{fig.S2} \end{center} \end{figure*}

Figure \ref{fig.pdfs} shows the volume-weighted distribution of density ($\rho$,
top row), speed
($v$, second row), magnetic field strength ($B$, third row) and gravitational potential
($\Phi$, bottom row) for each simulation at $t=0.$  For each plot, the solid black line shows the probability density
function (PDF) for the entire simulation for that quantity, which we denote
$V(q)$, where $q$ is $\rho, v, B$, or $\phi$.  The dashed black
line shows the distribution for just the preimage particles, which we denote
$V(q|*)V(*)$ (the PDF of $q$ given the fact that it will form a core.)  The two
black curves are measured directly from the simulations.  The solid grey
line shows the the ratio of the two distributions, $V(*|q)$, which shows the
probability of forming a core at a certain value of $q$.  The preimage PDF is somewhat non-intuitively denoted $V(q|*)V(*)$ (rather than just the first term) because the integral of this quantity gives $V(*)$, the probability that a particle ends in a core.

For density, $\rho$, and speed, $v$, we additionally have analytic forms for
$V(\rho|*)V(*)$ and $V(v|*)V(*)$ represented by the dashed grey lines.  
For
these, we additionally performed a
Kolmogorov Smirnov (KS) test to determine if the preimage distribution was drawn from
our analytic prediction.  The KS test compares the maximum difference, $D$, between the
cumulative distribution for two functions. It compares $D$ to the
critical value, $D_c$, and produces the $p$ value that they are drawn from the
same distribution.  All $p$ values are larger than 0.05, indicating that the
analytic prediction is a reasonable description of the preimage distribution.
Each will
be discussed in detail in the following sections.

For each simulation, $V(*)$ is the fraction of particles in preimage cores relative to the
total number, and is (0.057, 0.134, 0.13) for $(\sima,~\simb,~\simc)$
respectively.  It is also true that $V(*)=\int V(\rho|*)V(*) d\rho$, which we
will revisit in Section \ref{sec.discussion}.

We will discuss each distribution in turn.

\subsubsection{Density PDFs}
\label{sec.density_pdf}

It is hard to overstate the importance of the density PDF in modern star
formation theory.  It has been well established \citep{Vazquez-Semadeni94,
Federrath10, Collins12} that it can be
approximated by a lognormal, 
\begin{align}
    \label{eqn.lognormal}
    V(s) ds = \frac{1}{\sqrt{2\pi\sigmal^2}} 
    \exp{ 
    \frac{ -(s-\mu)^2}{2 \sigmal^2} ds
    }
\end{align}
where $s=\ln\rho/\rho_0$.
The actual distribution may have non-gaussianities that come from the divergence
of velocity in the driving
of the turbulence \citep{Federrath10}, intermittency \citep{Hopkins13},
magnetic fields \citep{Molina12}, and
complex thermodynamics \citep{Appel22},  to name a few.  The exact details of the
fit are not a major concern in this work, but the functional form is a useful
tool.

The top row of Figure \ref{fig.pdfs}, \ref{fig.pdfs}a, \ref{fig.pdfs}b, and
\ref{fig.pdfs}c, shows the PDF of density for our simulations, at the beginning
of the simulation.  The black line shows the PDF of density for the whole
computational domain, $V(\rho)$.  One can fit a lognormal to each one, but that does not concern us in this work.  This is a single
snapshot from a turbulent field, and as such, subject to random fluctuations that
cause it to deviate from lognormal.  What is of
interest is its relationship to the gas that actually forms stars.

The distribution of preimage gas is denoted by  $V(\rho|*) V(*)$ and can be seen as
the dashed black line in Figures \ref{fig.pdfs}a, \ref{fig.pdfs}b, and
\ref{fig.pdfs}c.  This is gas that contains
particles that we identify as ``forming stars'' at the end of the simulation
(see Section \ref{sec.identify} for our particle selection method.)  
Before this study, our assumption was that this $V(\rho|*)V(*)$ would be more-or-less a
step-function at some critical density, $\rho_c$.  This is not the case.  In
fact, the distribution is also roughly lognormal, with similar variance to the
total box, and a mean that is higher by a nontrivial amount.  
This distribution gives the gas that will form stars, so what is now needed is a
prediction of $V(\rho|*)V(*)$ that only depends on quantities available at the
beginning of the simulation and we will understand the star formation rate.  

We can predict $V(\rho|*)V(*)$ using Bayes' theorem.
We can measure the probability of a parcel of gas to become part of a core, given its
density, as
\begin{align}
	V(*|\rho) &= \frac{V(*) V(\rho|*)}{V(\rho)} \label{eqn.bayes}\\
    &= a_2 \rho^{a_1} \label{eqn.powerlaw}
\end{align}
Equation \ref{eqn.bayes} is Bayes theorem, and Equation \ref{eqn.powerlaw} is an empirical fit to the curve.  This is shown as the solid grey line
in Figures  \ref{fig.pdfs}a, \ref{fig.pdfs}b, and
\ref{fig.pdfs}c.  Thus, gas from all densities participate in the formation of
cores, with a powerlaw decline in the probability to low densities.  

An analytic description of the preimage PDF, $V(\rho|*)V(*)$, can then be found from Equation
\ref{eqn.bayes}:
\begin{align}
    V(\rho|*)V(*) &= V(*|\rho) V(\rho)\\
    &= a_2 \rho^{a_1} V(\rho).\label{eqn.predict}
\end{align}
This is shown as the dashed grey curve in the figures.  Here we can predict the
distribution of the preimage gas using only the initial PDF and knowledge of
$a_1$ and $a_2$.  The next task is to approximate $a_1$ and $a_2$, the slope and the normalization of
$V(*|\rho)$.  This becomes more speculative, so we relegate the discussion to
Section \ref{sec.discussion}.

We performed a Kolmogorov-Smirnov test between the measured preimage
distribution, $V(\rho|*)V(*)$, and the description $a_2 \rho^{a_1} V(\rho)$.  We
find that the two match with statistical significance of $p=(0.37, 1.0, 0.84)$
for the three simulations.  Since there are two fit parameters, good agreement
is not unexpected, but it demonstrates that the preimage gas is the product of a
powerlaw and a lognormal.  Or, as will be discussed in Section
\ref{sec.discussion}, another lognormal.


\subsubsection{Velocity PDFs}
\label{sec.velocity_pdf}

We show the velocity PDF in the second row, Figures 3d, 3e, and 3f.  As in the
top row, the black line shows the PDF of velocity for the initial snapshot of
the simulation, $V(v)$.  The black dashed curve shows the preimage velocity
distribution, $V(v|*)V(*)$.  The solid grey curve shows the probability of
forming a core at a given speed, $V(*|v)$.  The grey dashed line is our
prediction for the preimage distribution, $V(v|*)V(*)$.  We will discuss each of
these in turn.

The velocity PDF of a turbulent cloud, $V(v)$, should be very nearly a Maxwell-Boltzmann
distribution, $\mathcal{M}(v;\sigma)$, where
\begin{align}
    \mathcal{M}(v;\sigma) &= \frac{1}{\sqrt{2\pi \sigma^2}} v^2 \exp(-v^2/2
    \sigma^2).
\end{align}
This is for the same symmetry arguments that is used to derive it
in the classical theory of gases: to first approximation, the three components
$v_x, v_y$, and $v_z$, should be identical and separable, and the PDF should only
depend on the velocity magnitude \citep{Maxwell1860}.  This is enough to show that each velocity
component should be a Gaussian, and the total should be a Maxwell-Boltzmann
distribution with variance equal to the variance of any one component, i.e. the
1d Mach number.  See Rabatin and Collins (2023, in prep) for a more complete discussion.  
Thus, our volume-weighted velocity PDF is expected to be
\begin{align}
    V(v) &= \mathcal{M}(v;\sigmavone)
\end{align}
where $\sigmavone$ is the velocity variance along one dimension.
For our Mach number of 9, the 1d Mach number is
$\sigmavone\sim 9/\sqrt{3}\sim5.2$.  The measured $\sigmavone$ is found to be (5.2, 5.3, 5.4) for (\sima, \simb,
\simc).  

We find that the preimage gas fully samples the turbulent gas, with little
preference for low total velocity that one might expect.
For the preimage gas, we find that the simple prediction 
\begin{align}
    V(v|*)V(*) = V(*) \mathcal{M}(v;\sigmavone)
\end{align}
matches the data with a K.S. $p$ value of (0.87, 0.94, and 0.87) for the three
simulations, respectively.  This is
the dashed grey curve in Figures \ref{fig.pdfs}d, \ref{fig.pdfs}e, and
\ref{fig.pdfs}f, which lines up extremely well with the measured preimage PDF
shown by the black dashed line.  This is not a fit to the data.

To the extent that our prediction is correct, this implies
\begin{align}
    V(*|v) = V(*),
\end{align}
and the velocity gives us little to no information about the possibility for
collapse of a given parcel of gas.  This is not exactly correct, as the solid
grey curve is not completely flat in all three panels.  However, deviations from
a flat curve occur at very low and very high speed gas, which also suffer from
low number statistics as can be seen by the low values of $V(v)$.  There are no
trends in $V(*|v)$ that are consistent across all three simulations, which leads
us to conclude that the variations seen are due to the chaos of the turbulence.

In reality, the Maxwellian prescriptions are not perfect as we do
not really have the symmetry properties required for the derivation and we have
only presented a snapshot rather than an ensemble average.  However, the K-S
test between $V(v|*)V(*)$ and $\mathcal{M}(v;\sigmavone)$ give $p-$values of
$(0.87, 0.94, 0.87)$ for the probability that the two describe the same
distribution.  Thus a Maxwellian is a reasonable description of the preimage
speed distribution.

\subsubsection{Magnetic PDFs}
\label{sec.magnetic_pdf}

The PDF of magnetic field strength is shown in the third row, Figures
\ref{fig.pdfs}g, \ref{fig.pdfs}h,
and \ref{fig.pdfs}i.    The total PDF, $V(B)$, is shown in the black line; the
preimage magnetic distribution, $V(B|*)V(*)$, is dashed black; and $V(*|B)$ is
the solid grey curve.

Presently, we do not have a clear analytic form for $V(B)$.  We find that the distribution of the preimage gas is well described
by a fraction of the PDF for the whole gas:
\begin{align}
    V(B|*)V(*) = V(*) V(B)
\end{align}
and to the extent that this is true, $V(*|B) = V(*)$ and is constant with $B$.  There seems to be a small
excess of gas with high $B$ that forms stars. This could again be low number
statistics, or it could be correlation between $B$ and $\rho$, as high density gas has high magnetic fields, and we show in the first row that high density gas collapses preferentially.  A future paper
will examine the magnetic behavior in greater depth.

The preimage magnetic PDF can be seen to be a constant multiple of the parent.
Thus we compare $V(B|*)V(*)$, which is a measured quantity, with $V(*)V(B)$, a
constant multiple of the original distribution.  Thus we perform the KS test
between the two, and find $p=(0.30, 0.82, 0.41)$ that the preimage PDF is drawn
from a sample of $V(B)$.

\subsubsection{Potential PDFs}
\label{sec.potential_pdf}

There are only four independent fields in these simulations; density, velocity, magnetic
field, and gravitational potential.  For completeness, we also examine the last
of these.
The gravitational potential can be seen in the bottom row, Figures 3j, 3k, and
3l.  It is the most
beguiling of the distributions presented.  We lack a functional form for the
full $V(\phi)$, and the preimage distribution looks to be essentially unrelated.
There is a strong preference for gas that has low gravitational potential to
form cores.  As the potential is only set by the density field, and more
importantly the spatial distribution of the density, this shows that
the configuration of the density is more important than other quantities in
predicting if a parcel of gas will collapse.  

We hope to understand this further in the future.

\subsection{Density and Velocity structures}
\label{sec.means}
\subsubsection{Density Correlations}

Figure \ref{fig.means} shows a scatter plot of the mean density and rms velocity for each
preimage,
$\langle \rho/\rho_0\rangle$ and $\sigma_v/c_s$,  as well
as their histograms (left and top panels of that figure).    Here, the average is
taken over all the gas that makes up the preimage of an individual core.  Counter intuitively,
there are cores with mean densities quite low, lower than the mean of the box.
The typical mean density is about $2\rho_0$.  
The rms velocity is supersonic for all cores, with a
peak around $4 c_s$.

\subsubsection{Second Order Structure Function}

The second order longitudinal structure function, \sss, charachterizes how velocity scales
with size (or distance).  It is the average of the square of the velocity difference along the
separation:
\begin{align}
\sss &= \langle ( (\vvec(\xvec+\rvec)-\vvec(\xvec))\cdot
\hat{\rvec})^2\rangle_{\xvec, \hat{\rvec}}. \label{eqn.s2l}
\end{align}
That is, given two points separated by a vector $\rvec$, \sss\ is the average of
the velocity difference along $\rvec$.  
The average $\langle \rangle_{\xvec,\hat{\rvec}}$ is taken over all positions, $\xvec$,
and all directions, $\rhat$, leaving only the magnitude of the separation, $r$, as the free parameter.  
For incompressible turbulence, dimensional arguments yield $\sss = C
\epsilon^{2/3} r^{2/3}$, where $C$ is a (hopefully universal) constant and $\epsilon$ is the dissipation rate.
For our compressible simulations, we find $\sss\propto r^{a_2}$, where
$a_2=(0.76, 0.89, 0.91)$ for (\sima, \simb, \simc), respectively.  This can be
seen as the solid black lines in Figure \ref{fig.S2}.

For each preimage, we perform the same analysis but fix $\xvec$ on the center of each core.
By analogy to Equation \ref{eqn.s2l}, we plot, for each core,
\begin{align}
	S_{2,L,c_i} =\langle (\vvec - \vvec_{c_i}(r=0))\cdot
	\hat{\rvec})^2 \rangle_{c_i}
	\label{eqn.sigmavl}
\end{align}
where $c_i$ denotes a core, and $v_{c_i}(r=0)$ is the velocity at the center of the core.  Note that this is only an average over the
preimage zones, not including gas that does not end in the dense core. This is
similar to Equation \ref{eqn.s2l}, except that equation averages over center
position and angle, while Equation \ref{eqn.sigmavl} is at a fixed center.  

\def\meansl{\ensuremath{\langle S_{2,L,c_i} \rangle_{c_i}}}

Figure \ref{fig.S2} shows four things: \sss\ (black lines) computed for each simulation at
$t=0$; \sigmavl\ for each core in each simulation (morass of thin grey
lines); a heat map for \sigmavl\, showing the relative occupancy of each track in the space (color field); and
the average of $S_{2,L,c_i}$ over each core, \meansl\  (black dashed line).
Clearly there is a substantial spread in the behavior of the population of
cores, and we're taking a small sample that is more prone to sample variance.
\sss\ is a clearly defined powerlaw, while the average of $S_{2,L,c_i}$ is not so
regular.  However, $\sigmavl$ for each core clusters around \sss, and the average $\meansl$ is in the neighborhood of \sss, sharing slope and overall offset.  For $\sima$ and \simb, \meansl\ shares slope and offset with \sss.  The third simulation has a nontrivial offset between \meansl\ and \sss, indicating that preimages in \simc\ have slightly higher initial velocities than a typical patch of gas.

It should be noted that Figure \ref{fig.S2} is logarithmic, so the area above the black line is actually larger than the area below it, even though it appears the opposite.


\subsection{Length Scales}
\label{sec.length}

\begin{figure} \begin{center}
\includegraphics[width=\hw\textwidth]{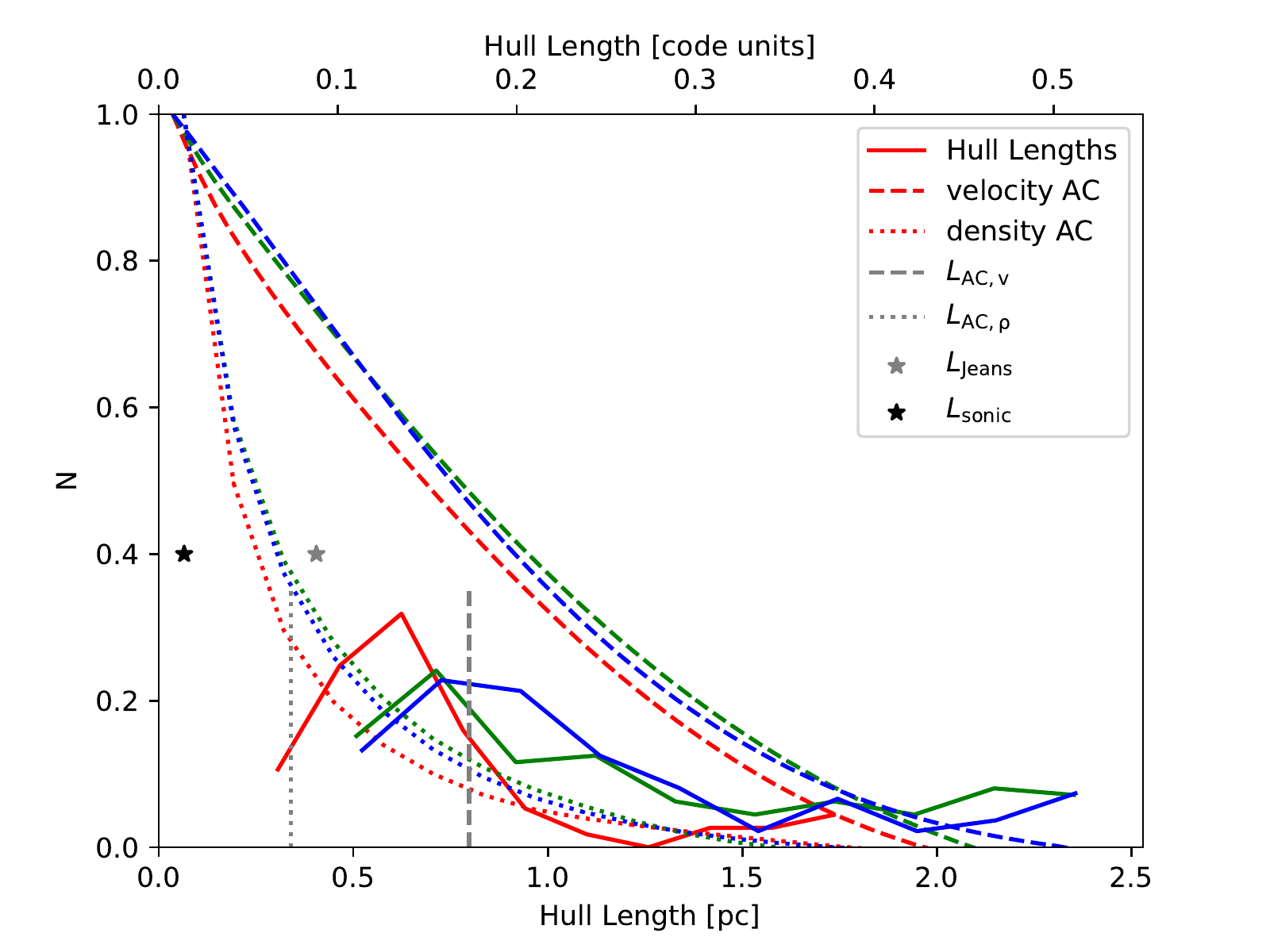}
\caption[ ]{Length Scales.  A histogram of preimage lengths (solid lines), density auto-correlation (dotted lines) and velocity auto-correlation (dashed lines).
The distribution of cores is in line with, but not exactly, the velocity auto-correlation length, $L_v$.  The density auto-correlation length (dotted vertical lines) is much shorter than the typical preimage size. }
\label{fig.lengths} \end{center} \end{figure}

Figure \ref{fig.lengths} shows the length scales at play in these simulations.
We compare the size scale of the \emph{preimage gas} (solid lines) to the size scales defined by
\emph{density structures} (dotted lines), scales defined by \emph{velocity
structures} (dashed lines), the
\emph{sonic length} (black star), and
the \emph{Jeans length} (grey star).  Each of these will be defined and discussed in turn.

We define the length scale of the preimage gas (solid lines in Figure
\ref{fig.lengths}) by way of the \emph{convex hull}.  
The convex hull defined by the points is the smallext convex polyhedron that
encloses the particles.  We will discuss these further in Section
\ref{sec.hulls}. For now
we simply use them as a quantifiable way to measure the length of the cloud of
preimage points.  The hulls can be seen in Figure \ref{fig.proj} as the black
lines around each set of colored points.  (In reality, the two-dimensional hull around the points is shown as the full three dimensional hull is difficult to visualize). The volume of the convex hull is easy to compute, and we take
the cube root of its volume to stand in for the length.  
Defining a single
length for an amorphous cloud of points is an oversimplification, the
oversimplification we select is to call $L_{\rm{preimage}}=V^{1/3}_{\rm{hull}}$.  
We find that the length
scales are broadly distributed, from 10\% - 50\% of the computational domain.
The strongly magnetized simulation is peaked around 0.14 $L_{\rm{box}}$, while the other two
are more broad and peak at roughly $0.2 L_{\rm{box}}$.  Thus one core requires (0.5-1.5) pc of molecular gas. 

Other estimates of the length are possible.
We could have taken the longest separation between points or the principle axis
of the moment of inertia tensor.  Each of these would result in a larger length
for our cores, since their cubes are larger than $V_{\rm{hull}}$.

The length scale of \emph{density structures} (dotted lines in Figure
\ref{fig.lengths}) is characterized by the density auto
correlation function:
\begin{align}
    AC_{\rm{\rho,3D}}(\delta \xvec) &= \frac{1}{V} \int_{V} d^3\xvec^\prime \rho(\xvec^\prime) 
    \rho( \xvec^\prime + \delta \xvec)\nn\\
    AC_{\rm{\rho}}(\delta x) &= \frac{1}{4 \pi \delta x^2} \int_{\Omega} d\Omega
    AC_{\rm{\rho,3D}}(\delta \xvec).
\end{align}
The first of these is the average of the product of density with itself shifted by $\delta
\xvec$.  The second is the azimuthal average of the first, making a function
only of the magnitude of the shift, $\delta x$.  This is plotted in Figure
\ref{fig.lengths} as colored dotted lines.  It is useful to further collapse the auto
correlation function to a single scalar characterizing the complex density
field.  This is the auto correlation length,
\begin{align}
    L_{\rm{AC},\rho} = \frac{\int AC_{\rho}(x) dx}{AC_{\rho}(0)},
\end{align}
and is shown as the grey dotted line.  As the lengths are similar for all
three simulations we plot their average.  All density hulls are larger than this
value, indicating that many density fluctuations feed a single core.  The
density structures that give rise to this $AC_\rho$ are determined by the
initial turbulence.

The \emph{velocity structures} (dashed lines in Figure \ref{fig.lengths}) are characterized by the velocity auto
correlation function.  This is defined similarly to that for density,
\begin{align}
    AC_{\rm{v,3D}}(\delta \xvec) &= \frac{1}{V} \int_{V} d^3\xvec^\prime
    \vvec(\xvec^\prime)  \cdot
    \vvec( \xvec^\prime + \delta \xvec)\nn\\
    AC_{\rm{v}}(\delta x) &= \frac{1}{4 \pi \delta x^2} \int_{\Omega} d\Omega
    AC_{\rm{v,3D}}(\delta \xvec) \\
    L_{\rm{AC,v}} &= \frac{\int  AC_{\rm{v}}(x) dx}{AC_{\rm{v}}(0)}.
\end{align}
The velocity auto correlation function and length are shown as dashed lines in
Figure \ref{fig.lengths}.  Colored dashed lines show the function, and the dark
grey dashed line shows $L_{\rm{AC,v}}$.  The peak of the distribution of core
lengths is smaller than this for the strongly magnetized simulation, and roughly
coincident with the peak for the other two.

\def\lsonic{\ensuremath{L_{\rm{sonic}}}}
For completeness, we also plot the \emph{sonic length} (black star) defined by way of
\begin{align}
    \sigma_v(L)^2 = c_s^2 (L/\lsonic)^p, \label{eqn.lsonic}
\end{align}
where $c_s$ is the speed of sound, and $p$ is expected to be about 1 for
supersonic turbulence.   To find $\lsonic$, we fit the
structure function (\sss, Equation \ref{eqn.s2l}, seen in the black lines in
Figure \ref{fig.S2}) to Equation \ref{eqn.lsonic}.  
We find that  $p=(0.76, 0.89, 0.91)$ for (\sima, \simb, \simc), and $\lsonic
\simeq 0.004$ for each.  
This is shown
as the black star in Figure \ref{fig.lengths}.  This is much
smaller than the length defined by the velocity auto correlation and the length
scale of the cores.

\def\ljeans{\ensuremath{L_{\rm{Jeans}}}}
Finally, again for completeness, we plot the Jeans length (grey star),
\begin{align}
    \ljeans=\frac{c_s}{\sqrt{G \rho_0}},
\end{align}
which is the largest disturbance that can be supported by pressure in uniform
gas.  This is shown as the grey star in Figure \ref{fig.lengths}.  It is similar
in scale to the density fluctuation scale, and much smaller than the objects
that do collapse.  It is coincidental that the Jeans length and density
autocorrelation lengths are comparable in size; the density autocorrelation
length is determined entirely by the turbulence before gravity is turned on, while the Jeans length depends on the strength of gravity.

We conclude that a typical
preimage blob is comprised of several density fluctuations, but only one or two
velocity fluctuations.  The Jeans length is smaller than the region that
collapses by a factor of 2.

\subsection{Convex Hulls and Spatial Overlap}
\label{sec.hulls}
\label{sec.overlap}

\begin{figure} \begin{center}
\includegraphics[width=\hw\textwidth]{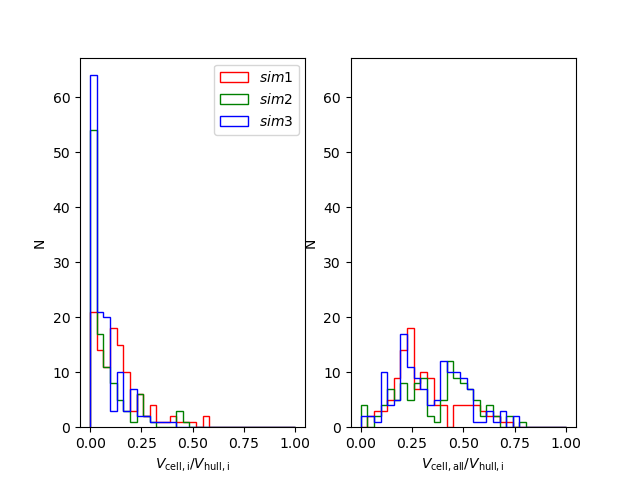}
\caption[ ]{(left) Volume filling fraction for each core relative to its
bounding hull.  Preimage gas is sparse in the domain.  (right) Volume filling fraction of
\emph{all cores} in each hull.  A large fraction of gas within a hull does not
make it to any core.}
\label{fig.volumes} \end{center} \end{figure}

\begin{figure*} \begin{center}
    \includegraphics[width=0.99\textwidth]{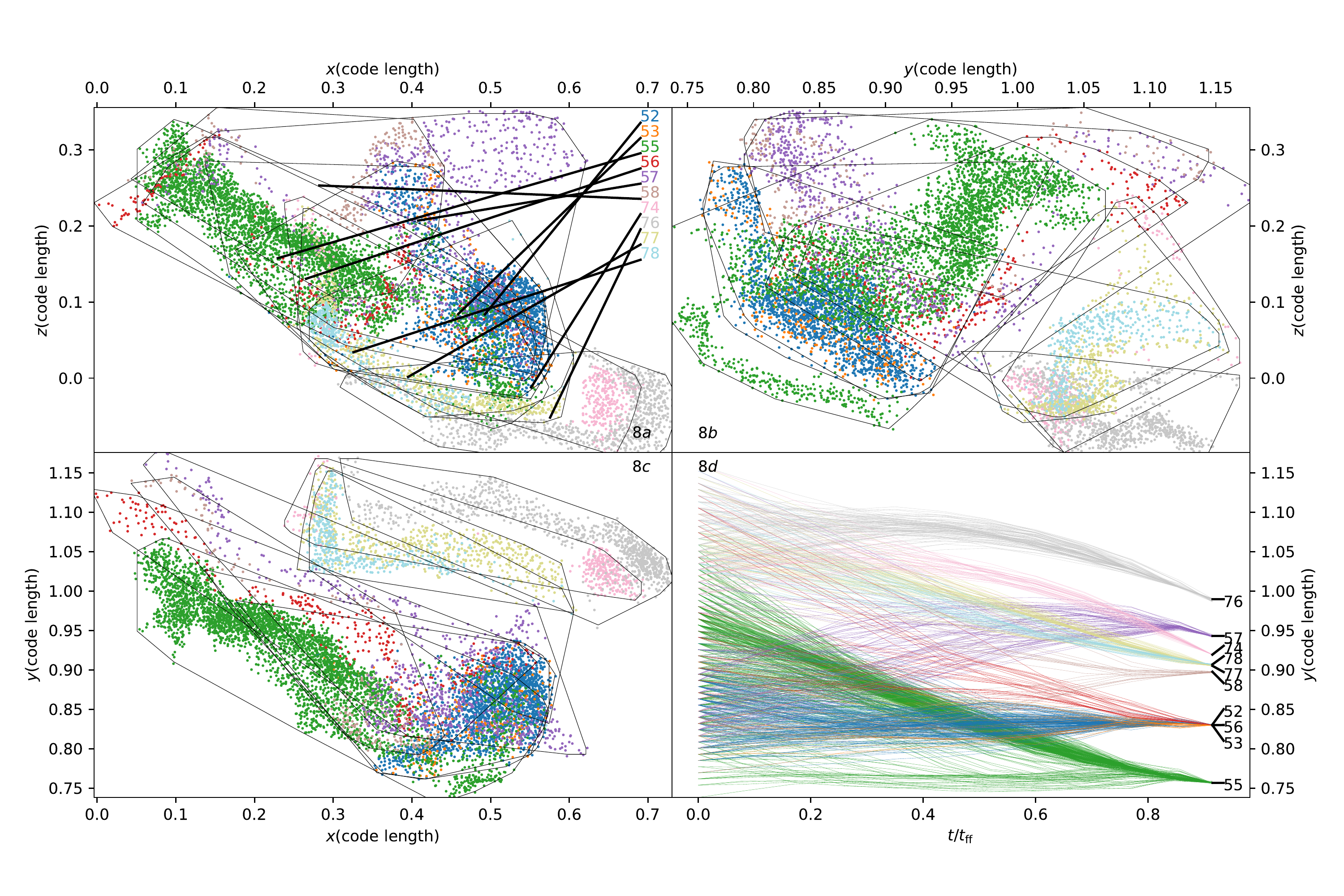}
\caption[ ]{\label{fig.overlaps} 
Three projections of a collection of preimage from \simc (8a-8c) and their time evolution
(8d).  Several preimages are quite
mixed with one another (e.g. 52,53,55, and 56) while some share boarders but do
not actually overlap (e.g. 74 and 76, in pink and grey).  The time evolution
shows that 74 and 76 separate early, while 52, 53 and 56 overlap for most of the
collapse, only separating at late times.
}
\label{} \end{center} \end{figure*}
\begin{figure*} \begin{center}
\includegraphics[width=0.32\textwidth]{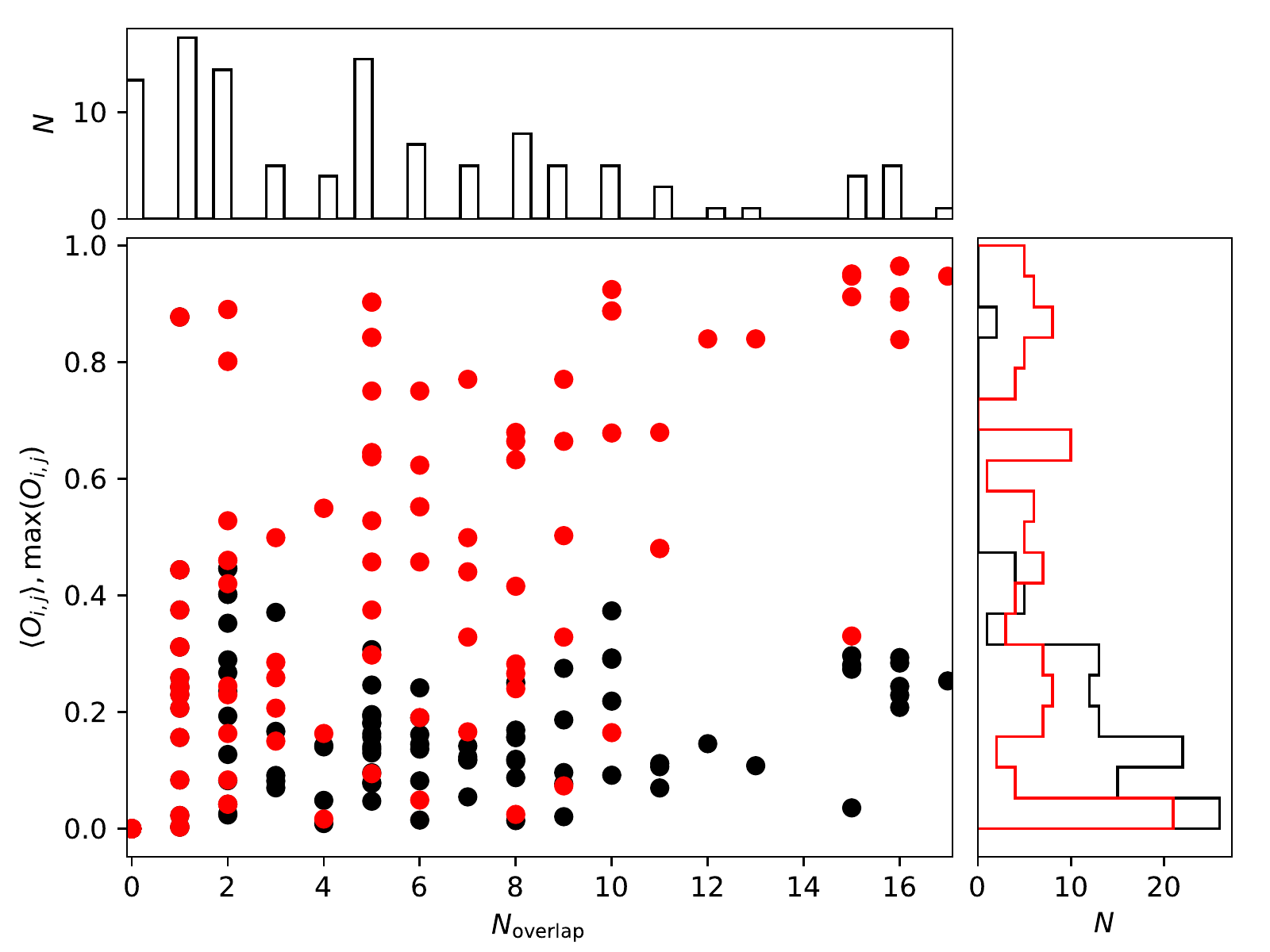}
\includegraphics[width=0.32\textwidth]{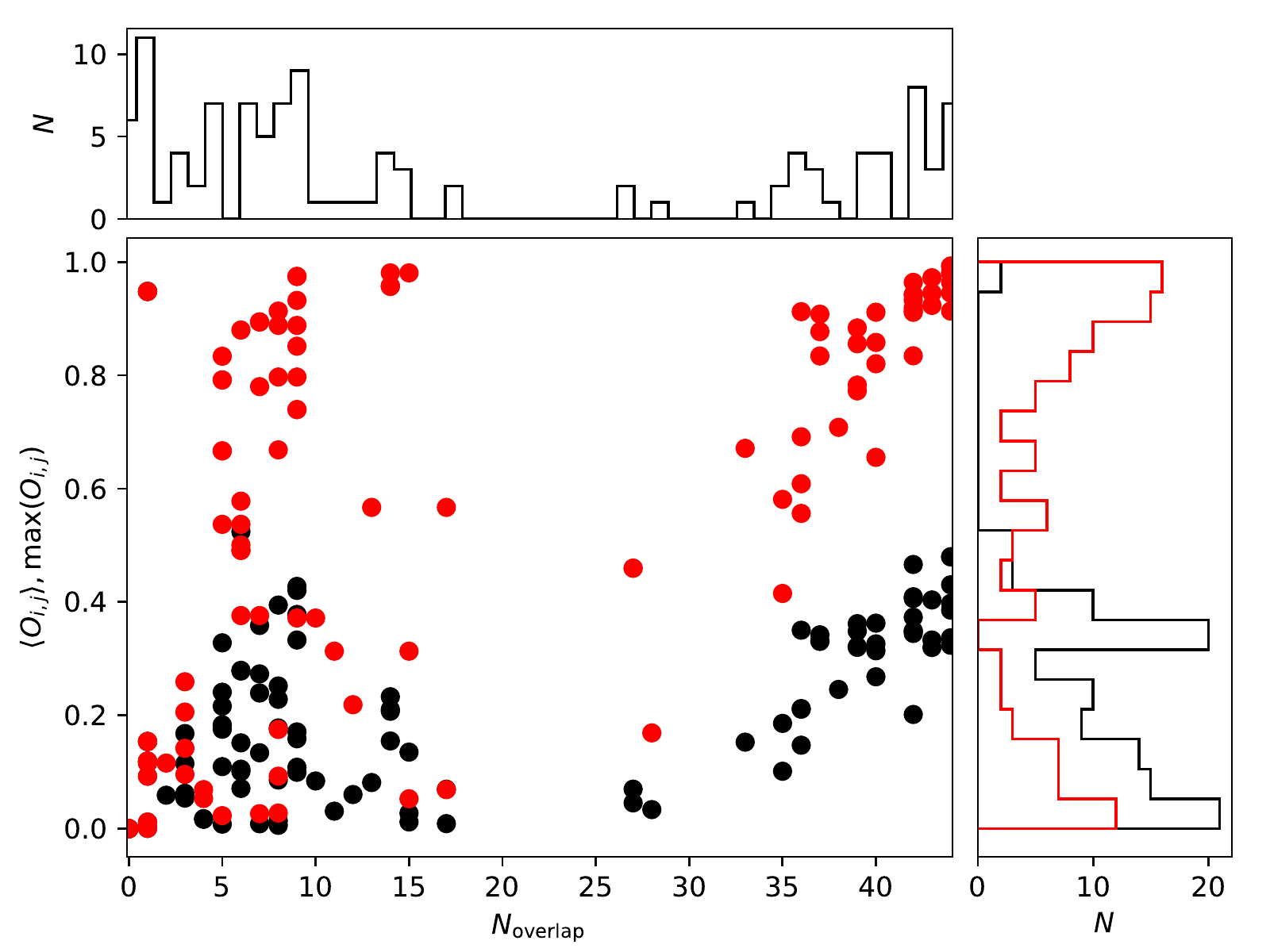}
\includegraphics[width=0.32\textwidth]{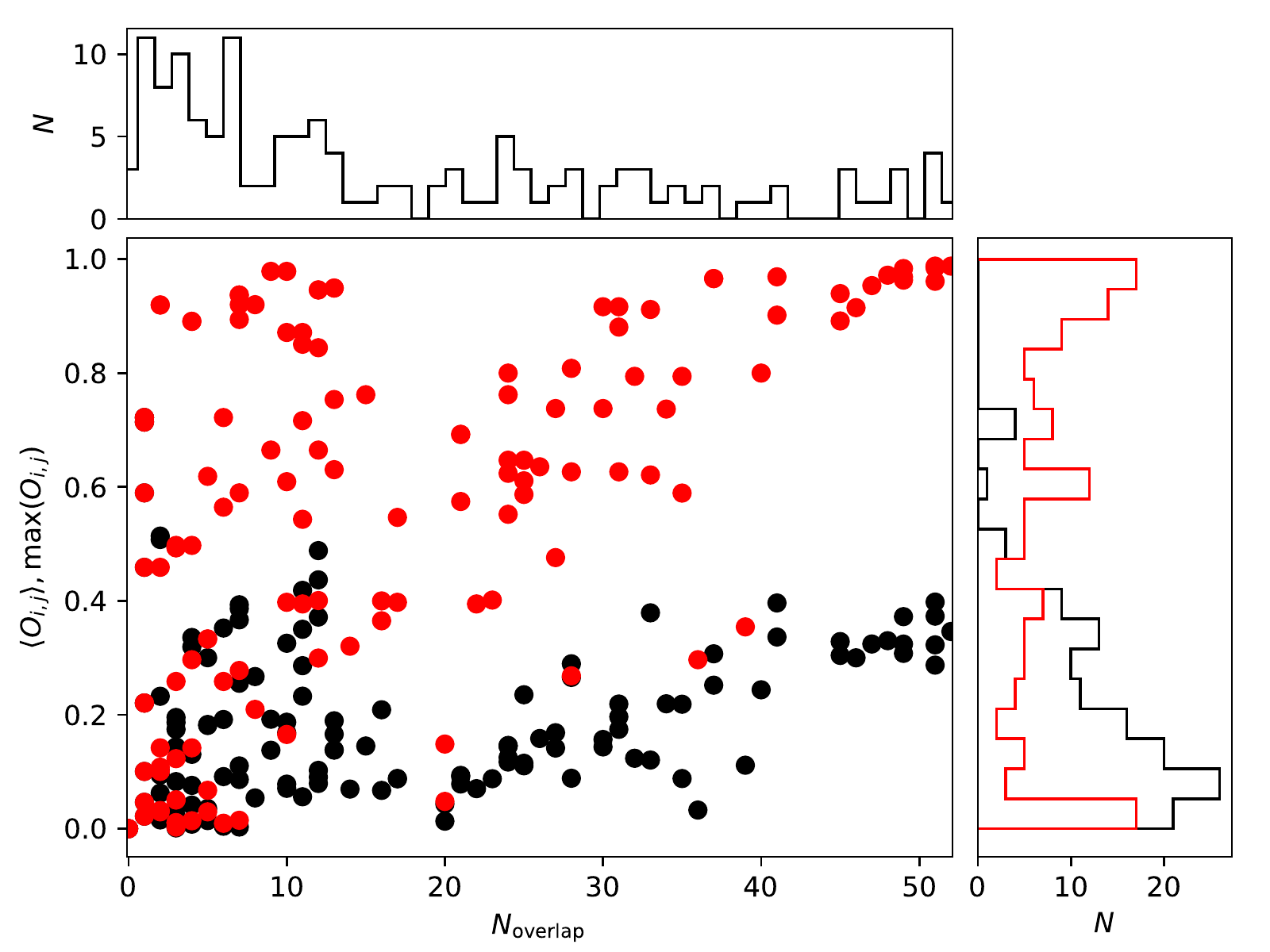}
\caption[ ]{Fraction and number of overlapping cores for each of our three
simulations.  For each figure, the center plot shows the average overlap fraction for each core (red) and the peak overlap fraction (black) vs. the number of strictly overlapping cores.  A few
points (notably at 0,0) represent more than one core.  Only (13, 6, 3) cores do
not overlap with any other cores in each simulation, respectively.  
}
\label{fig.overlap2} \end{center} \end{figure*}

\begin{figure*} \begin{center}
\includegraphics[width=0.32\textwidth]{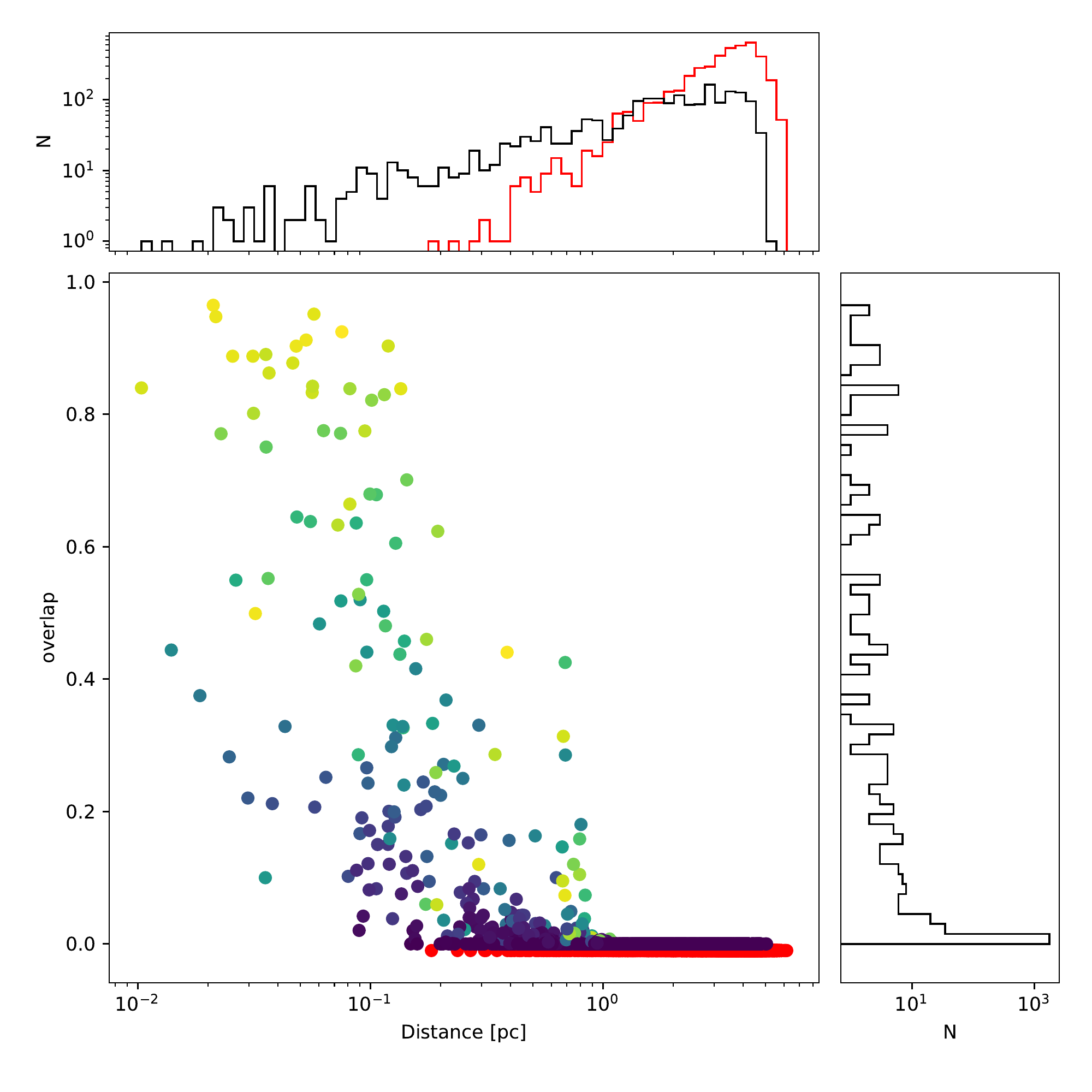}
\includegraphics[width=0.32\textwidth]{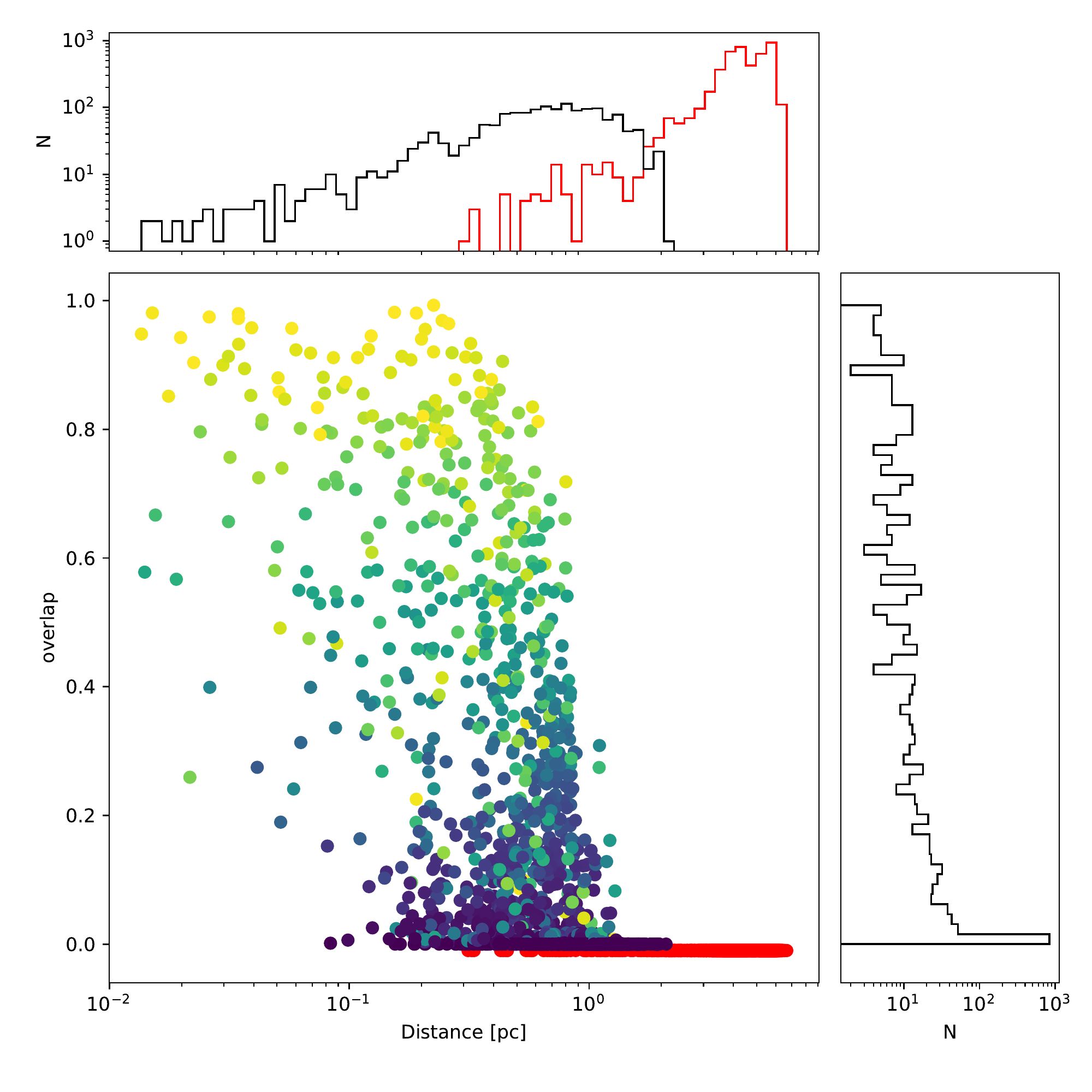}
\includegraphics[width=0.32\textwidth]{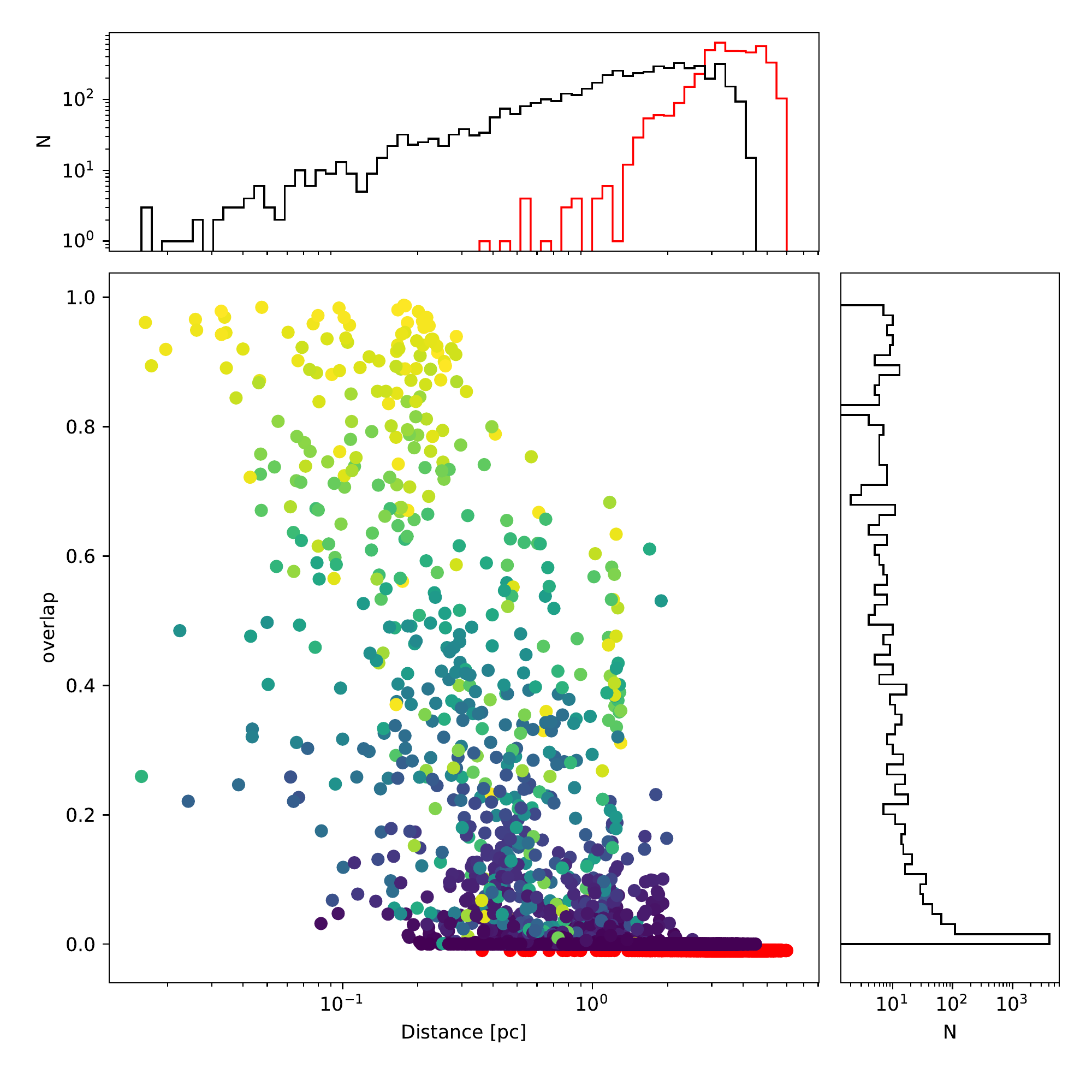}
\caption[ ]{Overlap Fraction vs. Final Binary Separation.  For every pair of
cores, the overlap of the two preimages is plotted vs. the final binary
distance.  Histograms showing the marginalized histograms are shown to the top
and right of each figure.  Red points come from pairs of cores with preimages
from different neighborhoods.  Color shows the ratio of overlaps between the cores,
with black being a ratio of 0 and yellow a ratio of 1.  Black points (low ratio) with
overlap > 0 have particles from core 1 are in the hull of core 2, but
particles from core 2 are not in the hull of core 1.  This implies apple-banana
nesting or complex abutting of the two cores.  All tight binaries come from gas that is originally mixed.
}
\label{fig.stay_close} \end{center} \end{figure*}

\begin{figure} \begin{center}
    \includegraphics[width=0.49\textwidth]{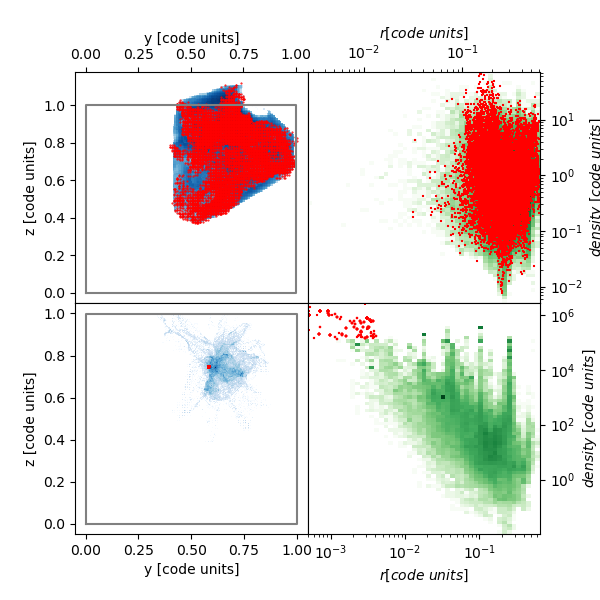}
\caption[ ]{The Other Ones.  Core particles (red) and particles within the
convex hull that do not end in cores (blue, green.)  The grey line denotes the
simulation boundary. The top row is the first snapshot and the bottom row is the final snapshot of the simulation.
Shown is core 107 from \simc, but most cores have similar
behavior.  \emph{Left}: spatial
location of core particles and ``other ones.''  \emph{Right}: Density-radius
relation for core particles (red) and other ones (green).}
\label{fig.otherones} \end{center} \end{figure}

\def\hull{\ensuremath{\mathcal{H}}}

In order to examine the spatial extents of the collapse, we draw the
\emph{convex hull} around the preimage gas for each of our cores.  The convex
hull was
introduced briefly in Section \ref{sec.length}, here we define it properly and
explore their contents.  

A surface,
 $\mathcal{C}$, is said to be \emph{convex} if for ever pair of points in
$\mathcal{C}$, the line segment joining the points is also in $\mathcal{C}$.
The \emph{convex hull} of a set of points, $\mathcal{P}$, is the smallest convex
surface that contains $\mathcal{P}$.  
For each core, $c_i$, we compute its convex hull $\hull_i$ by way of the Qhull
algorithm \citep{Barber96} as implemented in {\tt Scipy} \citep{Virtanen20}.

We plot the convex hull for the preimage
gas for all three simulations in the left column of Figure \ref{fig.proj}.  
There are two notable features; one is that the preimage points are decidedly
not convex, and the second is the high degree of spatial overlap between convex
hulls.  We discuss each of these in the next two sections.

\subsubsection{Convex Hull Filling Fractions}
\label{sec.volumefilling}

In any hull $\hull_i$ there are three populations of particles: particles that
end in core $c_i$, particles that end in a different core, $c_j$, and particles
that do not end in any core.  We define the volume of
the preimage gas, $V_{\rm{cell},i}$ for core $i$ as the total of the cell volume
of zones occupied by particles in core $i$.  The total volume of the hull
$V_{\rm{hull},i}$ is the volume of the polyhedron.
We discuss the third population in Section \ref{sec.others}.

Figure \ref{fig.volumes} shows the ratio of each hulls cell volume,
$V_{\rm{cell},i}$ to the hull volume, $V_{\rm{hull},i}$ in the left panel.  The
right panel shows the ratio of volumes for \emph{any} core to the hull,
$\frac{\sum_j V_{\rm{cell},j}}{V_{\rm{hull},i}}.$ For all three simulations this
distribution is peaked towards zero.  A large number of cores contain a small
number of particles that occupy a large volume of space.  No hull is more than
half filled by its own particles.  The take away from this is that preimage
particles are sparsely distributed in space.   When including particles that end
in other dense cores, the distribution of filling fractions is peaked at 25\%,
with a long tail to 75\%.  One region in space contains gas that will form
several cores.

There are two reasons for these low volume filling fractions.  The first is the mismatch between the hull and the "edge" of the preimage, like a banana in a plastic bag.  The second reason is the sparseness of the gas in space; most preimage zones in the volume touch zones that are not preimage zones.  The first is apparent from Figure \ref{fig.proj}.  The second is not observable in that figure due to projection effects, but we will revisit in Section \ref{sec.fractals}.

The reason for the substantial distinction between the left and right panels of
the volume filling distributions in Figure \ref{fig.volumes}, is the fact that the
preimage gas from different cores begin life spatially intertwined.  One hull can contain particles that go into many hulls.  The nature of this unmixing will be explored further in Paper II.

\subsubsection{Spatial Overlap of Convex Hulls}

The second most notable aspect about the convex hulls is the vast amount of
overlap between different preimages.  Quantifying this overlap is the primary
reason to use the convex hulls, as we can identify overlap by asking if
particles from one core are contained in the hull of another.  

Figure \ref{fig.overlaps} is a qualitative examination of this overlap and the
subsequent separation.  This set of cores can be seen in the bottom right of Figure
1g.  This figure shows three spatial projections and one space-time projection.
The spatial projections in panels 8a-8c are $\hat{y}, \hat{x}$, and $\hat{z}$, respectively and
the fourth panel is the $\hat{y}$ coordinate vs. time.  Each panel shares
an axis with its neighbors.  The black lines show the convex hull for each
preimage.  The primary purpose of this figure is to illustrate the extent of
spatial overlap between preimages.  Some are almost completely mixed
(e.g. 52,53,55,and 56) while some (74 and 76, in pink and grey respectively)
are not mixed, even though one appears wrapped around the other.  The
fourth panel, 8d, demonstrates the manner in which the cores untangle
themselves.  Core 55 is not assembled in one unit, but shows mergers of at least
two sub-clumps. It separates from 52, 53 and 56, which remain close until the
end.  Cores 74 and 76 separate quite early and end up not particularly
close; some as-of-yet unidentified parameter ensures their rapid separation.
Cores 77 and 78 are entirely mixed, and end up as a close binary.  

In order to quantify the amount of overlap each core experiences, we define the
overlap, $O_{i,j}$ between cores $i$ and $j$ as
\begin{align}
    f_{i,j} &= \frac{N_{i,j}}{n_i}\\
    O_{i,j} &= \min\left( f_{i,j}, f_{j,i} \right).
\end{align}
$N_{i,j}$ is the number of particle from core $i$ that are within the
convex hull of core $j$, and $n_i$ is the total number of particles in core $i$.
Thus $f_{i,j}$ is the fraction of $i's$ particles in hull $j$.  The overlap between the two cores, $O_{i,j}$, is then the smaller of $f_{i,j}$
and $f_{j,i}$.  
$O_{i,j} $ is
symmetric in $i,j$, but $N_{i,j}$ is not.
This definition is useful as it is only near unity when there is significant
mixing between the cores; for a pair of cores $i$ and $j$, a large $O_{i,j}$
means a large fraction of $i$ particles are in $j$ and vice versa.  In cases
where overlap occurs due to the non-convexity of one of the shapes, such as 74
and 76, a large fraction of 74 is with the hull of 76, but the converse is not
true, the particles of 76 lie outside of the hull of 74.  

Figure \ref{fig.overlap2} shows two points for every core.  The horizontal axis
shows the number of other cores each core overlaps with (i.e. $O_{i,j}>0$).  The vertical axis shows the
maximum overlap fraction (red) and average over all neighbors (black).  One can
see that there is a substantial degree of overlap between the cores.  For example, the first
panel shows \sima\, where $N_{\rm{overlap}}=16$, one finds three cores.  Each
of those cores overlaps with sixteen others.  For each, there is one other core
that overlaps at least 80\%. On average, each of the three cores overlaps with
its neighbors about 20\%.  For the other two simulations, even more overlap is
seen.  This is also evident in Figure \ref{fig.proj}.  As many as 40 cores
overlap in \simb, and 50 in \simc, and some of those overlap quite
substantially.

\subsubsection{Binaries Start Mixed}

It is interesting to ask what happens to gas that begins mixed.  Figure
\ref{fig.stay_close} shows, for every pair of cores, the overlap fraction vs.
the final binary separation distance.  The color shows the ratio of fractions,
$N_{i,j}/N_{j,i}$,
ranging from 0 to 1. By its definition, large values of $O_{i,j}$ will also have
ratios close to unity. Red points have no shared neighbors and all have $O_{i,j}=0$. Non-red points all have shared neighbors in common (i.e. friends-of-friends), even
if they themselves do not overlap.    
All close binaries in our simulation begin with mixed gas.

\subsubsection{The Other Ones}
\label{sec.others}
It is interesting to ask where the ``other'' particles go.  
Within the convex hull $\mathcal{H}_i$ there are three populations of
particles:   particles that go into the core $i$,   particles that go
into a different core, $j$,  and particles that do not go into any identified
core.  These ``other ones'' are shown in Figure \ref{fig.otherones} for core 107
from \simc.  Here, we
see two snapshots for core 107 of \simc;  the first snapshot (top row) and last
snapshot (bottom
row.) The left column shows the spatial distribution. The right column shows density vs.
radius from the center.   Red points are core particles. Blue (left) and green (right) show points that start within the convex hull but do not end on a core.  The ``other ones'' are
initially co-located with the core particles in both parameter spaces.  At the
end, the core particles are located in a few zones, at high density.  The
``other ones'' are found in a filamentary patch distributed in a large space,
around half the length of the simulation.  In the density-radius plot at late
time, it can be seen that the ``other ones'' are distributed in a
large volume, mostly at low density.  Several narrow spikes of population can be
seen, these are the location of other cores that form from this same volume.
The other cores themselves have also been cut out, they occupy much higher
densities than the ``other ones.'' 


This is just one representative core, but
others show similar trends.  Surprisingly enough (or perhaps not surprising), they are often spatially highly
filamentary.  
This plot can be seen for all cores in our core browser.  

Figure \ref{fig.otherones} 
also shows that a different particle selection method would not result in
qualitatively different results.  Much of the gas that starts in the hull ends
spread across much of the cloud.  Further, this gas will likely not ever become
part of the final population of stars, as radiation pressure will very soon
begin unbinding much of the gas that is close to the star.

\subsection{Fractals}
\label{sec.fractals}
 
\begin{figure*} \begin{center}
\includegraphics[width=\hw\textwidth]{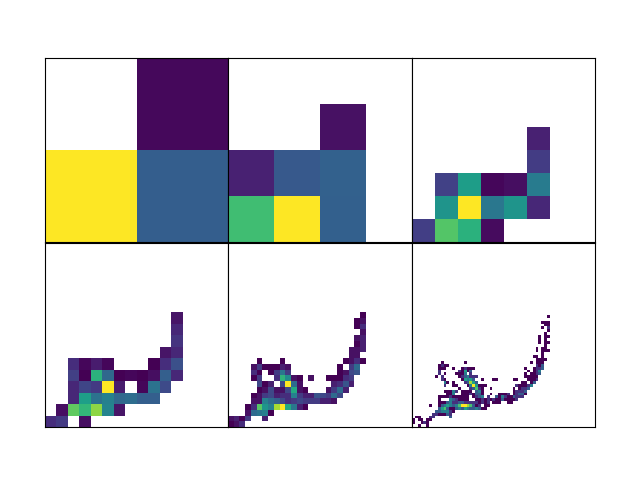}
\includegraphics[width=\hw\textwidth]{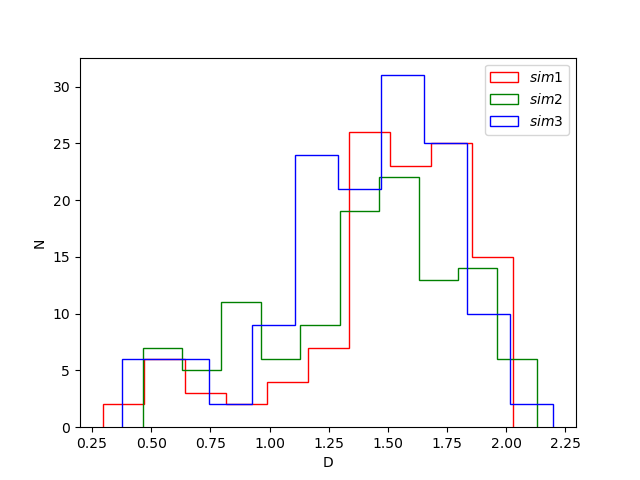}
\caption[ ]{Computing the Dimension of Cores.  \emph{Left:} Computing the
dimensionality of a preimage by box covering.  Here one core (\sima, core 323,
also seen in Figure \ref{fig.proj}a).  The scaling of the number of boxes
of length $\ell$ needed to cover the shape gives the dimension.  \emph{Right:}
The distribution of fractal dimensions for all three simulations.}
\label{fig.fractal} \end{center} \end{figure*}

Fractals are seen in many places in star forming clouds.  Here is another.  As anticipated in Section \ref{sec.volumefilling}, preimage zones within the convex hull are sparsely distributed in space.  We
use the Minkowski dimension, or box counting dimension, by counting the number, $N(\epsilon)$,
of boxes of size $\epsilon$ needed to cover the preimage, and measuring the
scaling as  $\epsilon$ becomes small.  The dimensionality of the preimage is defined
as
\begin{align}
    D = \lim_{\epsilon \to 0} \frac{\log N(\epsilon)}{\log 1/\epsilon}.
\end{align}
The volume of each particle is simply the volume of the zone it occupies.
  We cover each preimage with a uniform
grid of at most $128^3$, and count the number of boxes that contain particles.
We then double the box size and repeat.  The result is consistently a powerlaw,
and the exponent is the dimension.  

The distribution of dimensions can be seen in Figure \ref{fig.fractal}.  The left panel shows the box covering for core 323 of \sima, and the right shows the distribution of dimensions, $D$, for each core.
For all three simulations, the dimension peaks around 1.6.  This should come as
no surprise, as filamentary structures have been shown to resolve into smaller,
more filamentary structures upon improved observations.  

One of the shortcomings of using convex hulls to cover a fractal object is the
fact that it is impossible.  In fact, any 2d manifold cannot surround a
structure with fractal dimension less than 3 without also including some other
material.

\section{Discussion}
\label{sec.discussion}
\begin{figure*} \begin{center}
\includegraphics[width=0.32\textwidth]{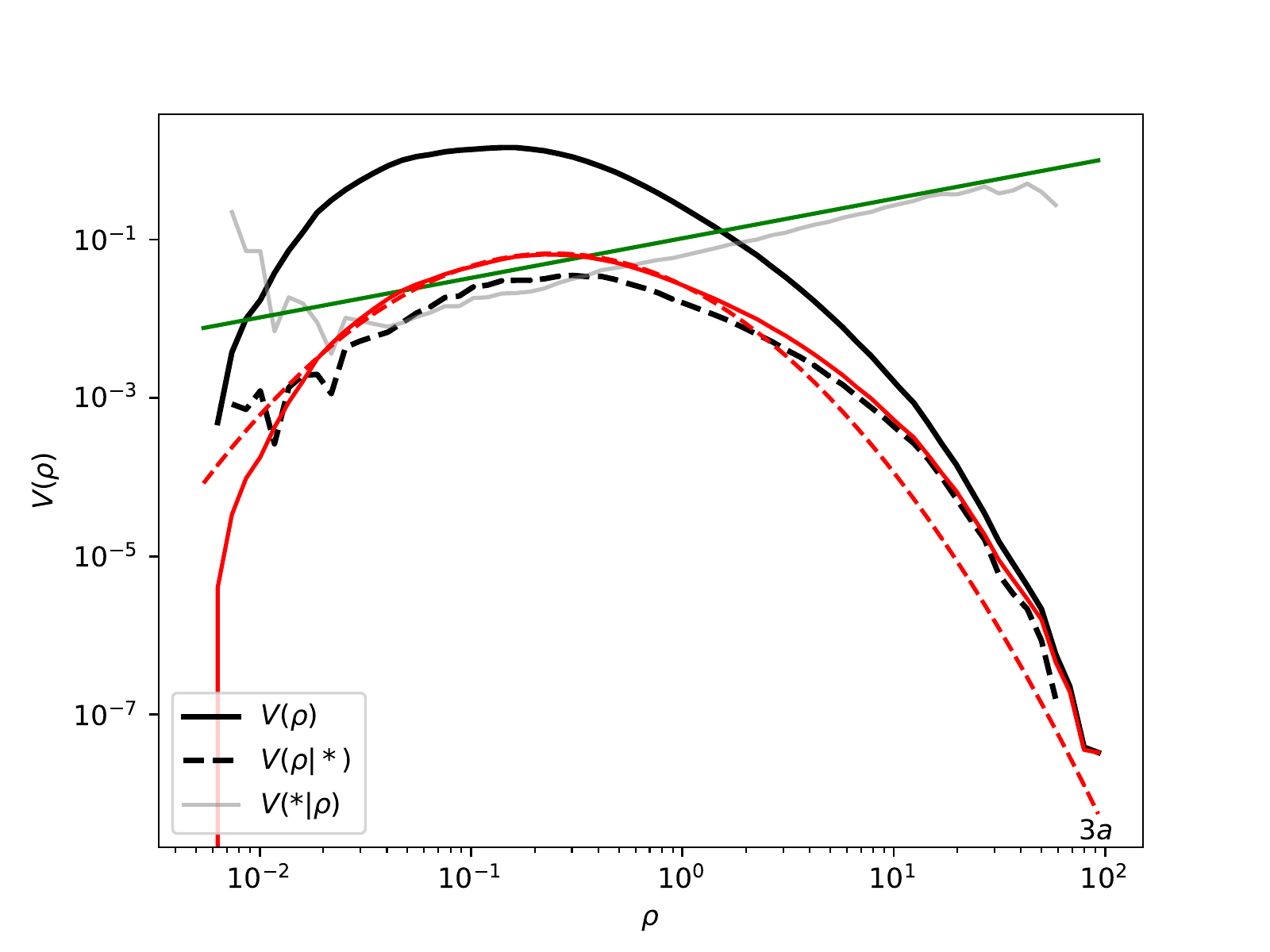}
\includegraphics[width=0.32\textwidth]{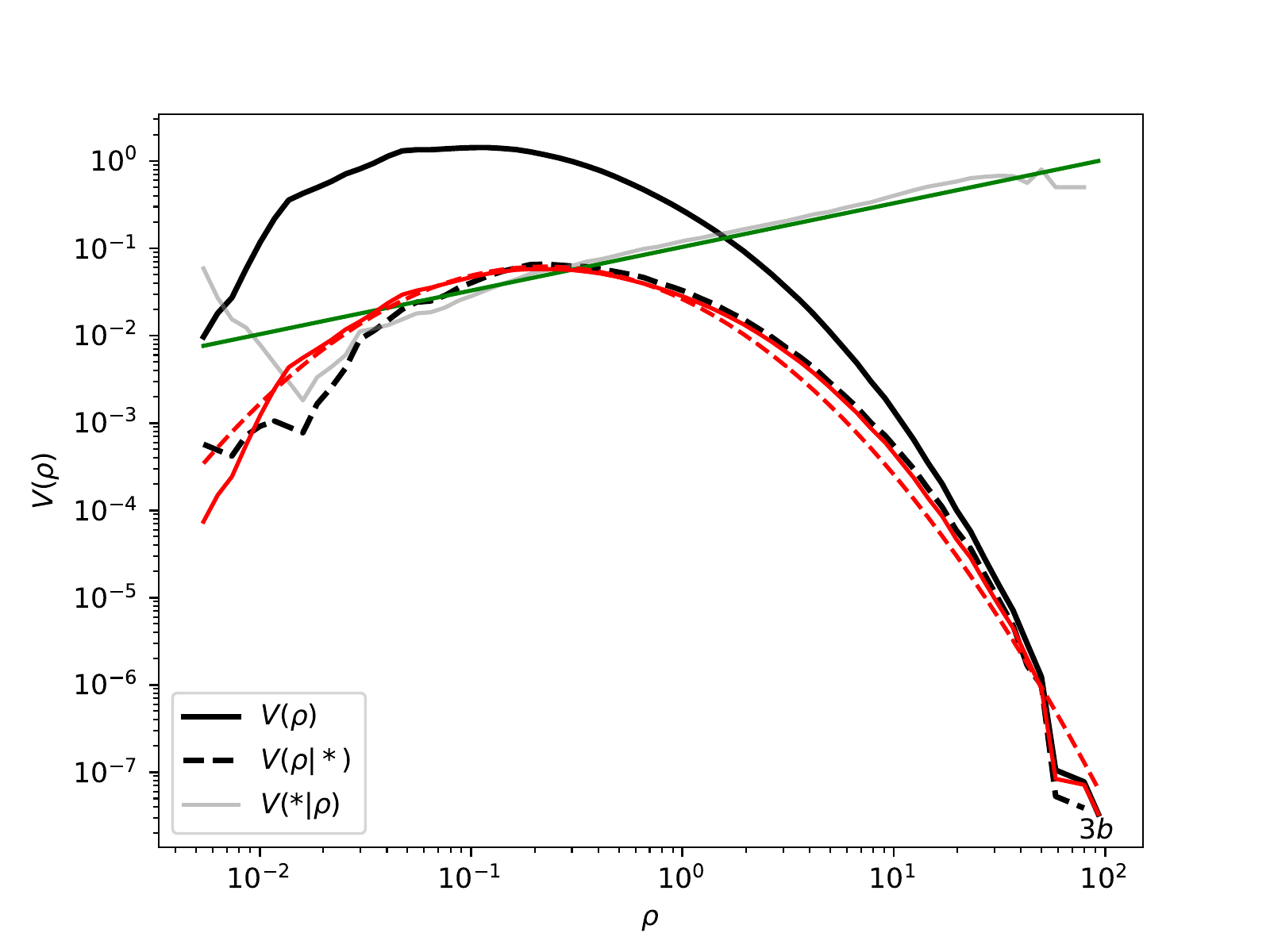}
\includegraphics[width=0.32\textwidth]{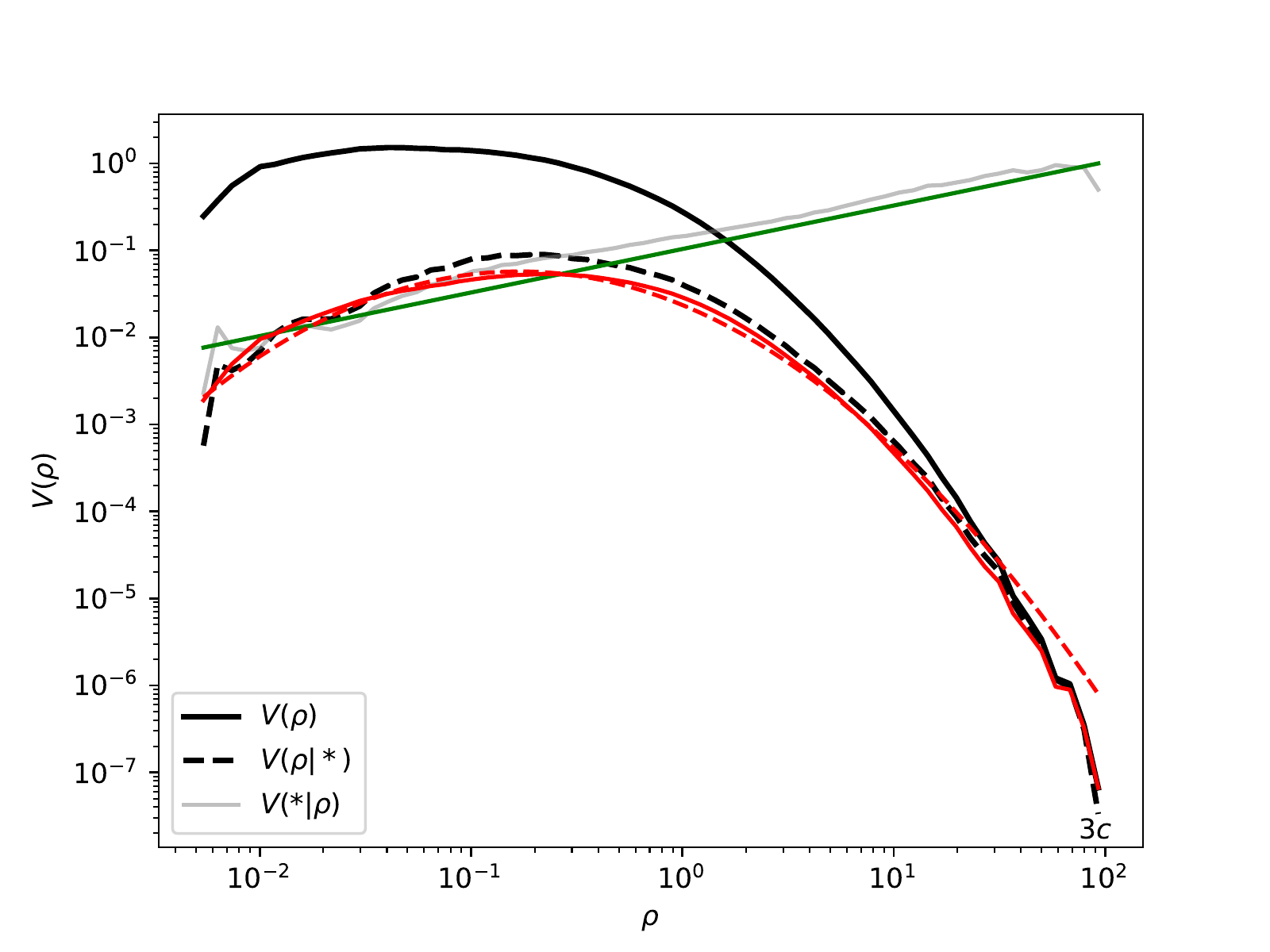}
\caption[ ]{The PDF of density with models.  As in Figure \ref{fig.pdfs}, the
black line is $V(\rho)$, dashed black is $V(\rho|*)V(*)$, grey is $V(*|\rho)$.
The green curve is an approximation to $V(*|\rho)$ given in Equation
\ref{eqn.suppress_a1}.  The solid red curve is the product of $V(\rho)$ and
$(\rho/\rhomax)^{1/2}$, and the dashed red curve is a lognormal prediction of
what will collapse based only on properties of $V(\rho)$. }
\label{fig.predict} \end{center} \end{figure*}

\subsection{Predicting the Star Formation Rate}
\label{sec.SFR}

We can use Figure \ref{fig.pdfs} to predict the star formation rate.  The prediction we present here is preliminary, and will be examined more fully in an upcoming study.  Here we take a probabilistic approach which has also been used elsewhere \citep{Elmegreen18}.

The integral under the dashed black line in Figures \ref{fig.pdfs}a,
\ref{fig.pdfs}b, and \ref{fig.pdfs}c gives the rate of core formation in our
simulation.  The simulation ran for one free fall time, and the number of
particles found in cores is
\begin{align}
    V(*) &= \int V(*|\rho) V(\rho) d\rho\\
    &\int V(\rho|*) V(*) d\rho.
\end{align}
The first line is integrating over the marginal distribution, and the
second is an application of Bayes theorem, ultimately showing that the
probability of forming a core is the integral under the dashed black line in Figures \ref{fig.pdfs}(a-c).
We have shown that $V(*|\rho)$ is well approximated by a powerlaw. If we can
describe that powerlaw by quantities easily obtainable from the beginning of the
simulation, we can predict the rate of star formation.  Which we shall now do.

The curve $V(*|\rho) = V(*) \rho^{a_1}$ has two properties that we will exploit
to make a predictive theory.  
First, $V(*|\rho) \simeq 1$ at $\rhomax$. All the densest gas forms cores. Second, the powerlaw slopes are all close to 1/2.
This is true for the current simulations, a future
study will explore the generality of these statements.  Thus the probability of
forming a star at a given density takes the simple form
\begin{align}
	V(*|\rho) &= a_2 \rho^{a_1} \label{eqn.predict_powerlaw}\\
    &\simeq \left(\frac{\rho}{\rhomax}\right)^{1/2}.
\end{align}
Inserting that into Bayes' theorem,
\begin{align}
    V(\rho|*) V(*) &= a_2 \rho^{a_1} V(\rho)\\
    &\simeq \left(\frac{\rho}{\rhomax}\right)^{1/2}
    V(\rho),\label{eqn.suppress_a1}\\
    &\simeq \frac{t_{\rm{ff}}(\rhomax)}{t_{\rm{ff}}(\rho)} V(\rho),
\end{align}
where we have used the fact that $t_{\rm{ff}}(\rho)=\left(G \rho \right)^{-1/2}$ in the last expression.  We can interpret this in the following manner:  all of the gas in the box
can participate in the collapse, but lower density gas is suppressed by its
ability to get dense.  It is probably an oversimplification to interpret this as ``low
density gas collapses slower'', but we shall make this statement.

We can further understand the mean and normalization of $V(\rho|*)V(*)$ by noticing that
the product of a powerlaw and a Gaussian is another Gaussian. By using
$\rho^a=e^{a \ln \rho}$:
\begin{align}
    \left(\frac{\rho}{\rhomax}\right)^a \exp{\left(\frac{-(s - \mu)^2}{2 \sigmal^2}\right)} &=
    A \exp{\left(\frac{-(s-(\mu+a
    \sigma^2))^2}{2
    \sigmal^2}\right)}\\
    A &= \exp{\left(\mu a + \frac{1}{2}a^2 \sigmal^2 - a s_{\rm{max}}\right)}.\label{eqn.gauss_predict}
\end{align}
So if $V(\rho)$ has mean $\mu$ and width $\sigmal$, $V(\rho|*)V(*)$ has width
$\sigmal$, mean $\mu+a\sigmal^2$, and is suppressed by $\exp{ \left(\mu a + a^2
\sigmal^2 - \frac{1}{2} a s_{\rm{max}} \right)}$. 

The solid red line in Figure \ref{fig.predict} is a prediction of the star formation
rate using only quantities available at the beginning of the simulation. Moreover,
it is a prediction of the distribution of gas that will form stars.  

In Figure \ref{fig.predict}, we show the applicability of these approximations.
As in Figure \ref{fig.pdfs}, the solid black line shows $V(\rho)$, the dashed black line shows
$V(\rho|*)V(*)$, and the grey line shows $V(*|\rho)$.  The green line shows an
approximation to $V(*|\rho)$ that does not require knowledge of the outcome of
the simulation, namely 
\begin{align}
    V(*|\rho) = \left( \frac{\rho}{\rhomax}\right)^{1/2}.
    \label{eqn.rhomax_powerlaw}
\end{align}
The solid red curve is the product of Equation
\ref{eqn.rhomax_powerlaw} and $V(\rho)$.  This is a reasonable match to the
black dashed curve it is trying to represent, but the error in the exponent on
the powerlaw shifts the peak of the prediction slightly.  Finally, we plot the
most aggressive oversimplification in the dashed red curve, which is the
Gaussian described by Equation \ref{eqn.gauss_predict}, with $\mu$ and $\sigmal$
and $s_{\rm{max}}$ taken from moments of $V(\rho)$.  
The advantage of the red curves is that we can write them down before the
simulation starts.

The long-term behavior of this model, specifically Equation
\ref{eqn.predict_powerlaw}, is of interest.  How would this change we were able
to continue the simulations and cores to accrete more?
Likely it will not evolve much with continued
accretion, as the gas near a given core generally came from the same convex
hull, and will sample the same initial distribution.  The distribution
$V(q|*)V(*)$ samples the distributions of gas in the union of the convex hulls
of all the cores, further accretion will simply sample this gas better.
However, it is also possible that ``low density gas collapses slower'' is not an
oversimplification but is instead accurate, in which case the powerlaw will
become more shallow with time.  It may depend on the manner in which accretion
is finally halted by radiation.

We indicate that $V(*|\rho)$ is relatively insensitive to mean magnetic field,
but our simulations were performed with a single Mach number and cloud mass.  A
future study with more simulations will address the validity of this model as Mach number and cloud mass are changed.

\section{Conclusions}
\label{sec.conclusions}
In this work, we embedded pseudo-Lagrangian tracer particles in an Eulerian
adaptive mesh refinement simulation of a collapsing molecular cloud.  This is
the first paper in a series; in this first paper, we focus on the properties of the
initial conditions of the gas before it collapses.    The \emph{preimage} gas
refers to 
the gas at the beginning of the simulation that contains tracers that are found
in prestellar cores at the end of the simulation.

The most salient feature is that the preimage gas forms a wide variety of
morphologies.  Many are isolated single or small multiple systems that travel
most of the length of the box before the end of the simulation.  Many form from
converging flows.  Many form a major structure early and accrete in both clumpy
and continuous manners.  Many form along filaments.  A quick glance at Figure
\ref{fig.hair} demonstrates this, and a lengthy visit to our online browser will
show this in abundance.

The preimage gas bears substantial imprint of the turbulence from which it was
born, but the gravity is what ultimately dictates the collapse.

We measure the probability distribution function (PDF) for density, $\rho$,
speed, $v$, magnetic field, $B$, and gravitational potential, $\phi$.  We measure
$V(q)$, the PDF of the quantity $q$ at $t=0$, as well as $V(*|q)$, the
probability the gas will form a star at that value of $q$.  For density, we find
that $V(\rho)$ is roughly lognormal as expected, and $V(*|\rho)$ is a powerlaw in density.
This shows that high density gas is more likely to form cores, but gas at all
densities can participate in the collapse.  The speed PDF, $V(v)$, is roughly
Maxwellian, with width set by the one-dimensional Mach number, as expected.  We
find that $V(*|v)$ is basically flat, with small deviation caused by the small
statistics of a single turbulent field.  This indicates that the velocity at the
start of the collapse is not a strong predictor of what gas will collapse.  The
magnetic PDF, $V(B)$, has not known functional form, more importantly $V(*|B)$
is also basically flat.  There is perhaps a small increase in more highly
magnetized gas, but this is likely a result of that gas also being denser, and
denser gas preferentially collapses.  The potential PDF, $V(\phi)$, also has not
known form, but is shown to be strongly predictive of what gas will collapse, as
$V(*|\phi)$ is a linearly decreasing function of $\phi.$  As expected, gas
in deeper potential wells collapses.  

We also examine the 2nd order structure function of velocity for both the
individual preimages and the full volume.   The structure
function for the full volume is a powerlaw in the separation, as expected. 
Each preimage is a single instance of a turbulent field, but taken as an
ensemble they are reasonably well described by the structure function of the
whole volume.  

We examine the length scales of the preimage gas, and compare to other length
scales in the simulation.  Preimage gas is large in space, many times larger
than the density auto correlation length, and one to a few times larger than the
velocity auto correlation length.  A given preimage typically covers many
density fluctuations and a few velocity fluctuations.  

We show that the preimage gas is sparse in space.  We measure the volume filling
fraction and show that it is small compared to the overall size of the preimage.
We also compute the fractal dimension for each preimage, and find that preimages
have a fractal dimension that peaks around 1.6.

We show that preimage gas for different cores begins mixed in space.  Close
binaries in the end of the simulation come from gas that is initially well
mixed.

Finally, we present a predictive model for the star formation rate.  This is
based on the finding that $V(*|\rho)$ is a powerlaw that has a value of unity at
the highest density and an index of roughly 1/2.  We then interpret $V(*|\rho)$
as suppression of collapse based on the free-fall time; denser gas can collapse
faster.  This has a similar functional form as other predictions of the star
formation rate, but a different interpretation.  This prediction will be
explored in a future study.

\section*{Data Availability}
Simulation data used in this work are available upon reasonable request
(dccollins@fsu.edu).

\section*{Acknowledgements}

Support for this work was provided in part by the National Science Foundation
under Grant AAG-1616026.  Simulations were performed on \emph{Stampede2}, part
of the Extreme Science and Engineering Discovery Environment
\citep[XSEDE;][]{Towns14}, which is supported by National Science Foundation grant number
ACI-1548562, under XSEDE allocation TG-AST140008.

\bibliographystyle{mnras}
\bibliography{apj-jour,ms}  

\end{document}